%% file: MAIN.tex
\documentclass[journal]{IEEEtran}
\input{usepackage}

\newacronym{agv}{AGV}{Automated Guided Vehicle}
\newacronym{ai}{AI}{Artificial Intelligence}
\newacronym{iot}{IoT}{Internet of Things}
\newacronym{nbiot}{NB-IoT}{narrowband Internet of Things}
\newacronym{nr}{NR}{New Radio}
\newacronym{ntn}{NTN}{Non-Terrestrial Networks}
\newacronym{leo}{LEO}{Low-Earth Orbit}
\newacronym{meo}{MEO}{Medium-Earth Orbit}
\newacronym{geo}{GEO}{Geo-synchronous Earth Orbit}
\newacronym{cn}{CN}{Core Network}
\newacronym{dlt}{DLT}{Distributed Ledger Technology}
\newacronym{e2e}{E2E}{End-2-End}
\newacronym{isl}{ISL}{Inter-Satellite Link}
\newacronym{gnb}{gNB}{Next Generation NodeB}
\newacronym{aoi}{AoI}{Age of Information}
\newacronym{ue}{UE}{User Equipment}
\newacronym{pdcch}{PDCCH}{Physical Downlink Control Channel}
\newacronym{prach}{PRACH}{Physical Random Access Channel}
\newacronym{nprach}{NPRACH}{Narrowband \gls{prach}}
\newacronym{npdcch}{NPDCCH}{Narrowband \gls{pdcch}}
\newacronym{pusch}{PUSCH}{Physical Uplink Shared Channel}
\newacronym{pdsch}{PDSCH}{Physical Downlink Shared Channel}
\newacronym{poi}{PoI}{Proof-of-Inclusion}
\newacronym{pow}{PoW}{Proof-of-Work}
\newacronym{ra}{RA}{Random Access}
\newacronym{rao}{RAO}{Random Access Opportunity}
\newacronym{dl}{DL}{Downlink}
\newacronym{ul}{UL}{Uplink}
\newacronym{gs}{GS}{Ground Station}
\newacronym{5gc}{5GC}{5G Core Network}
\newacronym{upf}{UPF}{User Plane Function}
\newacronym{ngso}{NGSO}{Non-Geostationary Orbit}
\newacronym{rtt}{RTT}{Round-Trip Time}
\newacronym{tmtc}{TMTC}{Telemetry and Telecontrol}
\newacronym{pmf}{pmf}{Probability Mass Function}
\newacronym{qos}{QoS}{Quality of Service}
\newacronym{fcfs}{FCFS}{First Come First Serve}
\newacronym{sinr}{SINR}{Signal-to-Interference-and-Noise Ratio}
\newacronym{harq}{HARQ}{Hybrid Automatic Repeat Request}
\newacronym{embb}{eMBB}{Enhanced Mobile Broad-Band}
\newacronym{mmtc}{mMTC}{Massive Machine-Type Communication}
\newacronym{urllc}{URLLC}{Ultra-Reliable Low-Latency Communication}
\newacronym{rrc}{RRC}{Radio Resource Control}
\newacronym{fl}{FL}{Federated Learning}
\newacronym{ml}{ML}{Machine Learning}
\newacronym{nn}{NN}{Neural Network}
\newacronym{lpwa}{LPWA}{Low-Power Wide Area}
\newacronym{iiote}{iIoTe}{Intelligent IoT Environment}

\newacronym{gd}{GD}{Gradient Descent}
\newacronym{lag}{LAG}{Lazy Aggregated Gradient Descent}
\newacronym{admm}{ADMM}{Alternative Direction Method of Multipliers}
\newacronym{gadmm}{GADMM}{Group \gls{admm}}
\newacronym{dgadmm}{D-GADMM}{Dynamic \gls{gadmm}}
\newacronym{ggadmm}{GGADMM}{Generalized  \gls{gadmm}}
\newacronym{cadmm}{C-ADMM}{Censored  \gls{admm}}
\newacronym{cggadmm}{C-GGADMM}{Censored  \gls{ggadmm}}
\newacronym{cqggadmm}{CQ-GGADMM}{Censored  Quantized  \gls{ggadmm}}

\begin{document}

\title{\huge Learning, Computing, and Trustworthiness in Intelligent IoT Environments: Performance-Energy Tradeoffs}

\author{ 
Beatriz Soret, \textit{IEEE Member}, 
\thanks{Beatriz Soret, Lam D. Nguyen, and Petar Popovski are with the Connectivity Section, Department of Electronic System, Aalborg University, Denmark. \textit{Contact: \{bsa, ndl, petarp\}@es.aau.dk}.}
Lam D. Nguyen, \textit{IEEE Graduate Student Member},
Jan Seeger,
\thanks{Jan Seeger, Arne Br\"{o}ring, and Vivek Kulkarni are with Siemens AG, Munich, Germany. Contact: jan.seeger@thenybble.de, \textit{\{arne.broering, vivekkulkarni\} @siemens.com}.}
Arne Br\"{o}ring,
Chaouki Ben Issaid, \textit{IEEE Member},
\thanks{Chaouki Ben Issaid, Sumudu Samarakoon, Anis El Gabli, and Mehdi Bennis are with Centre for Wireless Communications (CWC), University of Oulu, Finland. \textit{Contact: \{Chaouki.BenIssaid, Sumudu.Samarakoon, Anis.Elgabli, mehdi.bennis\} @oulu.fi  } }
Sumudu Samarakoon, \textit{IEEE Member}, 
Anis El Gabli, \textit{IEEE Member}, \\
Vivek Kulkarni,
Mehdi Bennis, \textit{IEEE Fellow},  
and 
Petar Popovski, \textit{IEEE Fellow}
}

\markboth{IEEE Transaction on Green Communication and Networking}%
{Shell \MakeLowercase{\textit{et al.}}: Bare Demo of IEEEtran.cls for IEEE Journals}

\maketitle

\begin{abstract}
An \gls{iiote} is comprised of heterogeneous devices that can collaboratively execute semi-autonomous IoT applications, examples of which include highly automated manufacturing cells or autonomously interacting harvesting machines. Energy efficiency is key in such edge environments, since they are often based on an infrastructure that consists of wireless and battery-run devices, e.g., e-tractors, drones, \gls{agv}s and robots. The total energy consumption draws contributions from multiple \gls{iiote} technologies that enable edge computing and communication, distributed learning, as well as distributed ledgers and smart contracts. This paper provides a state-of-the-art overview of these technologies and illustrates their functionality and performance, with special attention to the tradeoff among resources, latency, privacy and energy consumption. Finally, the paper provides a vision for integrating these enabling technologies in energy-efficient \gls{iiote} and a roadmap to address the open research challenges. 
\end{abstract}

\glsresetall


\IEEEpeerreviewmaketitle

\input{1-introduction}

\input{2-background}

\input{4-enabling-technologies}

\input{5-towards-energy-efficient-intelligent-iot}

\input{6-conclusions}

\section*{Acknowledgment}

This work has received funding from the European Union’s Horizon 2020 research and innovation programme under grant agreement No. 957218 (Project IntellIoT).

\bibliographystyle{ieeetr}
\bibliography{MAIN}


\begin{IEEEbiography}[{\includegraphics[width=0.9in,height=1.55in,clip,keepaspectratio]{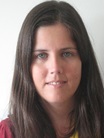}}] {Beatriz Soret}[M’11] received her M.Sc. and Ph.D. degrees in Telecommunications from the University of Malaga, Spain, in 2002 and 2010, respectively. She is currently an associate professor at the Department of Electronic Systems, Aalborg University, and a Senior Research Fellow at the Communications Engineering Department, University of Malaga. Her research interests include LEO satellite communications, distributed and intelligent IoT, timing in communications, and 5G and post-5G systems.
\end{IEEEbiography}

\begin{IEEEbiography}[{\includegraphics[width=0.9in,height=1.55in,clip,keepaspectratio]{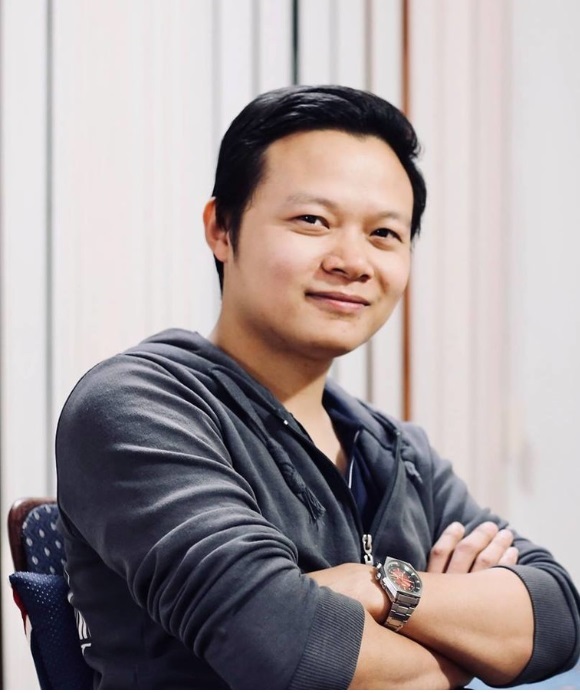}}] {Lam D. Nguyen}
(S'20) is a Ph.D. Fellow at Aalborg University. He obtained Master Degree in Computer Science at Seoul National University, and a Bachelor in Telecommunication at Hanoi University of Science and Technologies in 2019 and 2015, respectively. His research includes Distributed Systems, Blockchain, Smart Contracts, the Internet of Things, and applying Blockchain and Federated Learning to enhance the efficiency of Blockchain-based IoT monitoring Networks. He receives Outstanding Paper Award for the research about scaling Blockchain in Massive IoT at the IEEE World Forum Internet of Things 2020, travel grant from Linux Foundation 2020, Best Research Award for a solution of Blockchain-based CO2 Emission Trading from VEHITS 2021. He is Hyperledger Member, IEEE Student Member, and IEEE ComSoc Student Member. 
\end{IEEEbiography}

\begin{IEEEbiography}[{\includegraphics[width=0.9in,height=1.55in,clip,keepaspectratio]{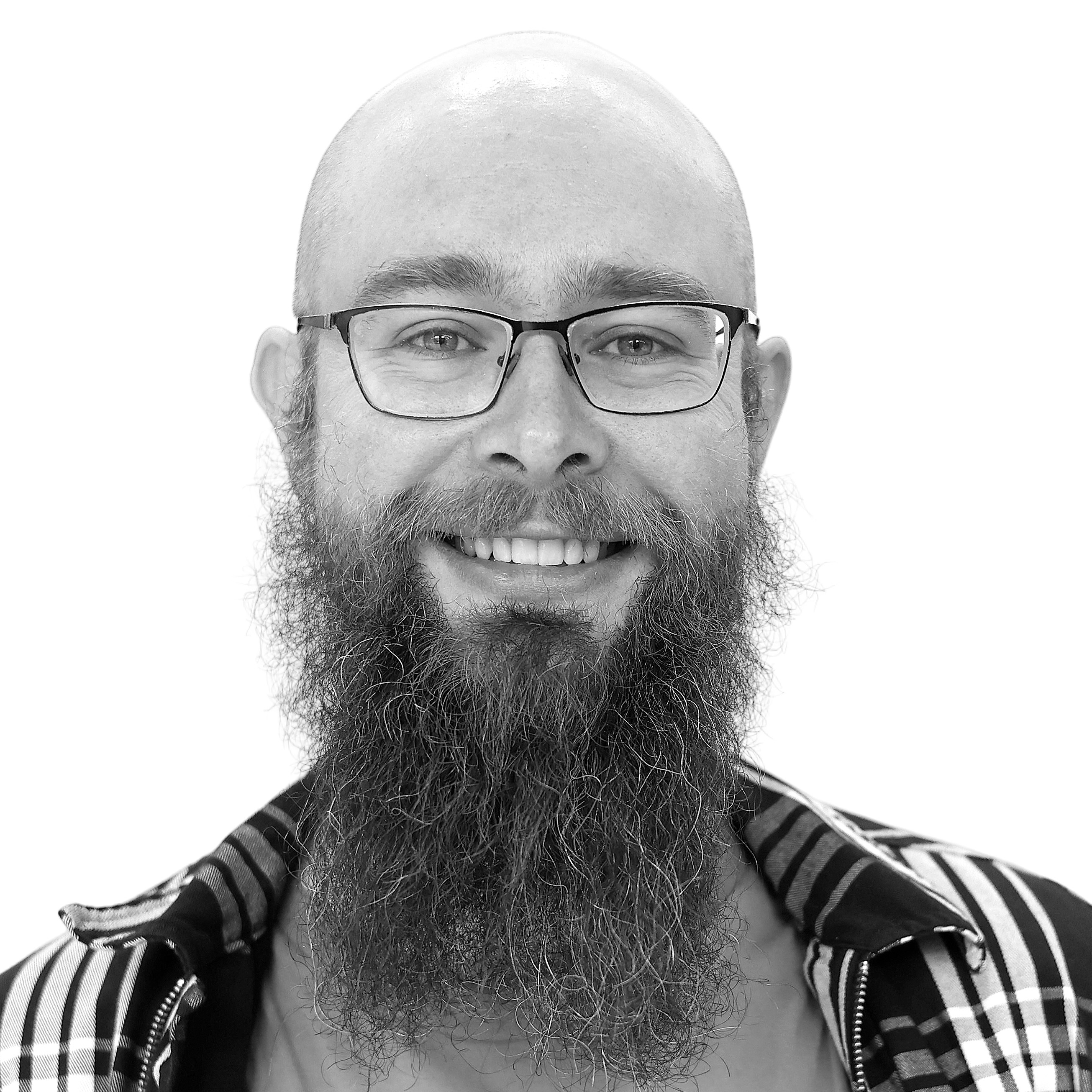}}] {Jan Seeger}
is a Ph.D. Researcher with the Technical University of Munich, Munich, Germany. He is active in the areas of Internet of Things and automation research,
and how semantic technologies can improve the engineering of automation systems. He received the
M.Sc. degree in computer science from the TU Munich. 
\end{IEEEbiography}

\begin{IEEEbiography}[{\includegraphics[width=0.9in,height=1.55in,clip,keepaspectratio]{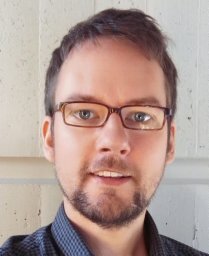}}] {Arne Br\"{o}ring}
is a Senior Key Expert Research Scientist at Siemens Technology in Munich. He received his PhD in 2012 from the University of Twente (Netherlands). Dr. Br\"{o}ring has contributed to over 90 publications in the field of distributed systems and has served on various program committees and editorial boards. His research interests range from distributed system designs, over sensor networks, and Semantic Web, to the Internet of Things. At Siemens, he has been in charge of the technical \& scientific coordination of large EU research projects (BIG IoT and IntellIoT). Before joining Siemens, Dr. Br\"{o}ring worked for the Environmental Systems Research Institute in Zurich, the 52°North Open Source Initiative, and led the Sensor Web and Simulation Lab at the University of Münster.
\end{IEEEbiography}

\begin{IEEEbiography}[{\includegraphics[width=0.9in,height=1.55in,clip,keepaspectratio]{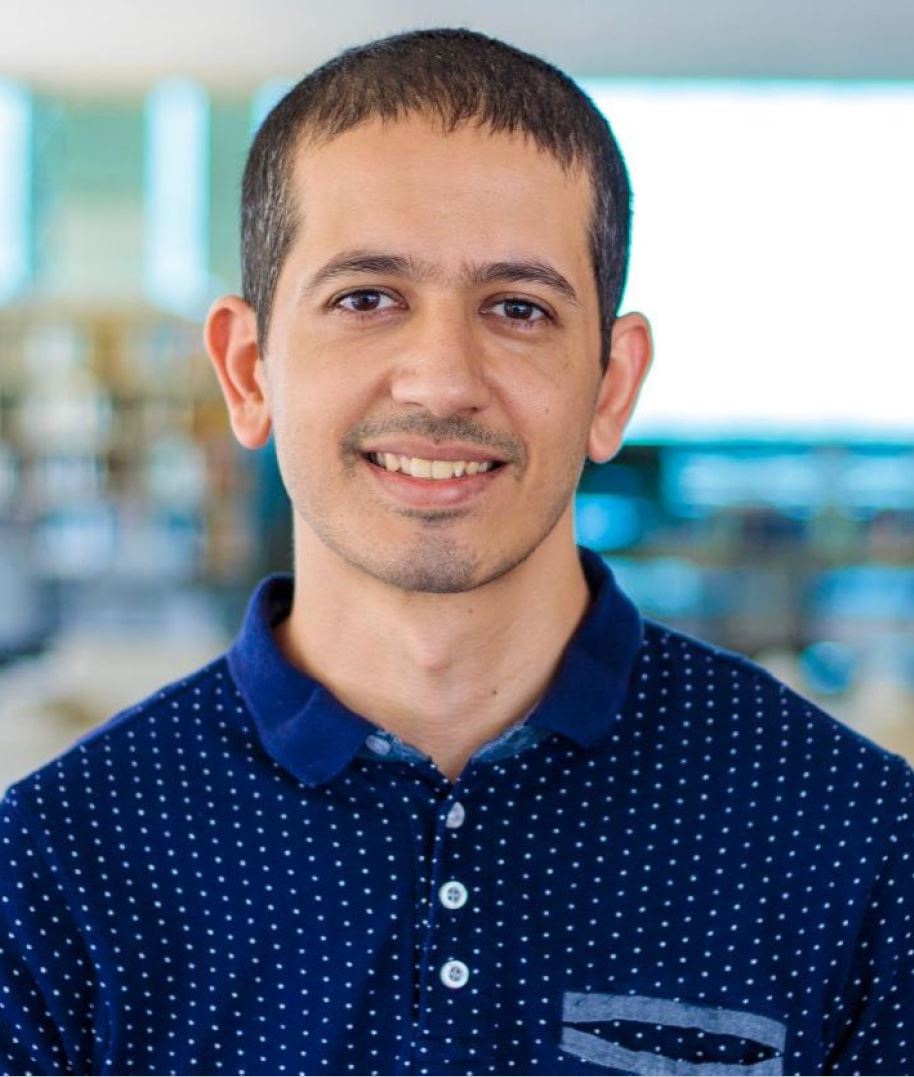}}] {Chaouki Ben Issaid} received the Dipl\^ome d'Ing\'enieur with majors in economics and financial engineering from the Ecole Polytechnique de Tunisie (EPT) in 2013, the master's degree in applied mathematics and computational science (AMCS) and the Ph.D. degree in statistics from King Abdullah University of Science and Technology (KAUST) in 2015 and 2019, respectively. He is currently a Post-Doctoral Fellow with the Centre for Wireless Communications (CWC), University of Oulu. His current research interests include communication-efficient distributed learning and machine learning applications for wireless communication.
\end{IEEEbiography}

\begin{IEEEbiography}[{\includegraphics[width=0.9in,height=1.55in,clip,keepaspectratio]{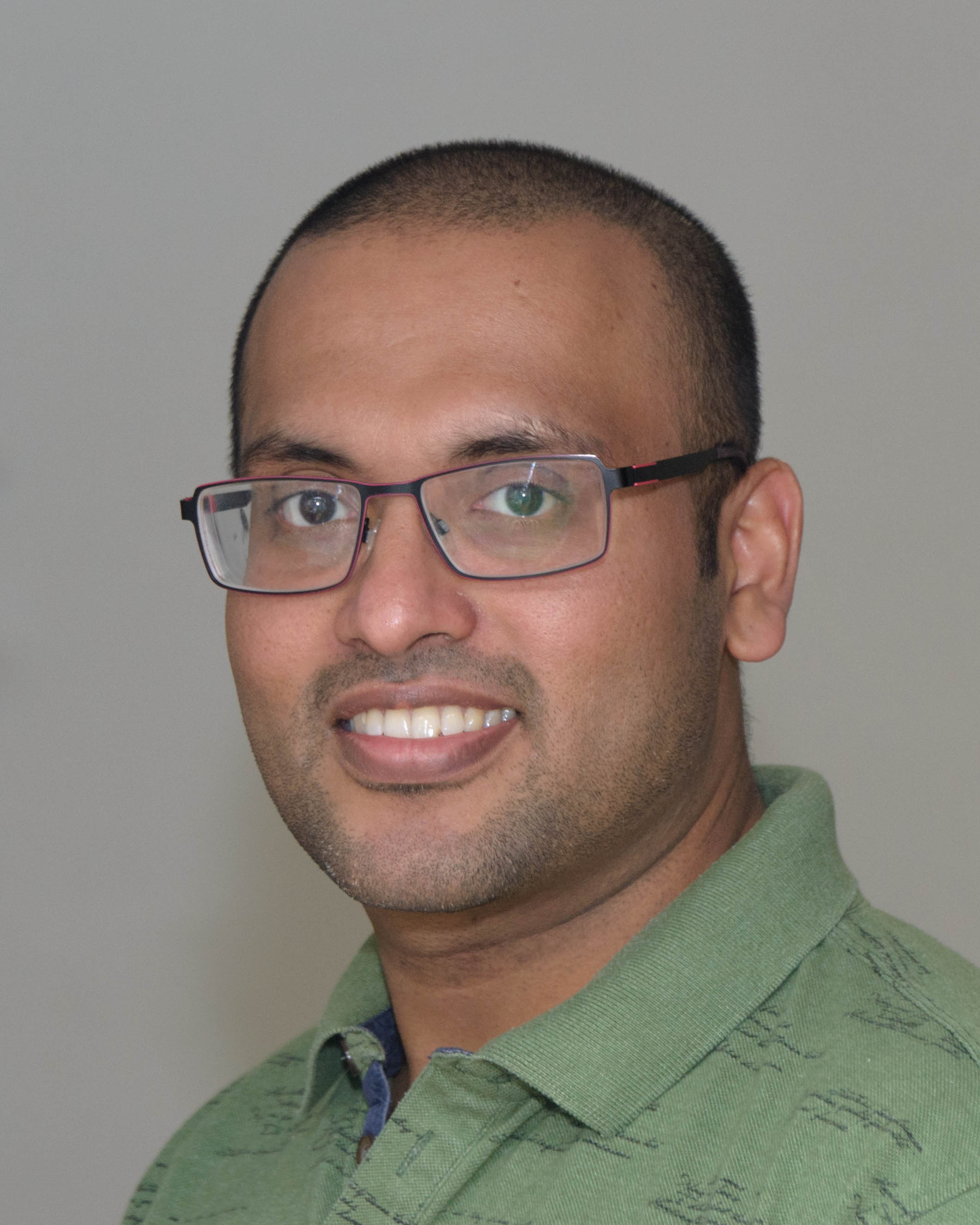}}] {Sumudu Samarakoon}[M'12]
received the B.Sc. degree (honors) in electronic and telecommunication engineering from the University of Moratuwa, Sri Lanka, in 2009, the M.Eng. degree from the Asian Institute of Technology, Thailand, in 2011, and the Ph.D. degree in communication engineering from the University of Oulu, Finland, in 2017. He is currently an Assistant Professor and a member of intelligent connectivity and networks/systems (ICON) group with the Centre for Wireless Communications, University of Oulu. His main research interests are in heterogeneous networks, small cells, radio resource management, machine learning at wireless edge, and game theory. Dr. Samarakoon received the Best Paper Award at the European Wireless Conference, Excellence Awards for innovators and the outstanding doctoral student in the Radio Technology Unit, CWC, University of Oulu, in 2016. He is also a Guest Editor of Telecom (MDPI) special issue on “millimeter wave communications and networking in 5G and beyond”. 

\end{IEEEbiography}

\begin{IEEEbiography}[{\includegraphics[width=0.9in,height=1.55in,clip,keepaspectratio]{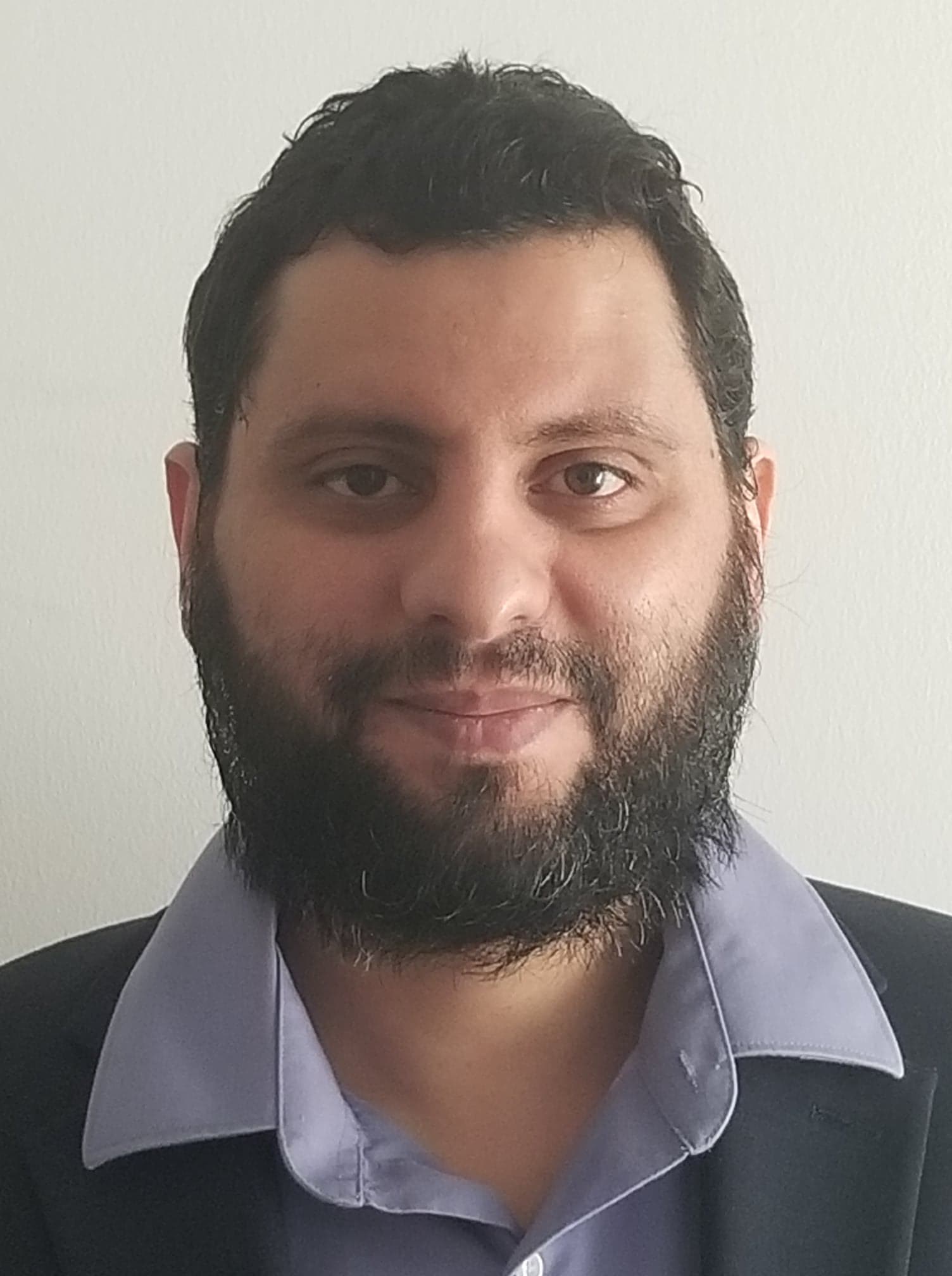}}]{Anis Elgabli} received the B.Sc. degree in electrical and electronic engineering from the University of Tripoli, Libya, in 2004, the M.Eng. degree from UKM, Malaysia, in 2007, and the M.Sc. and Ph.D. degrees from the Department of Electrical and Computer Engineering, Purdue University, West Lafayette, IN, USA, in 2015 and 2018, respectively. He is currently a Post-Doctoral Researcher at the Centre of Wireless Communications, University of Oulu. His main research interests include heterogeneous networks, radio resource management, vehicular communications, video streaming, and distributed/decentralized machine learning and optimization. 
\end{IEEEbiography}

\begin{IEEEbiography}[{\includegraphics[width=0.9in,height=1.55in,clip,keepaspectratio]{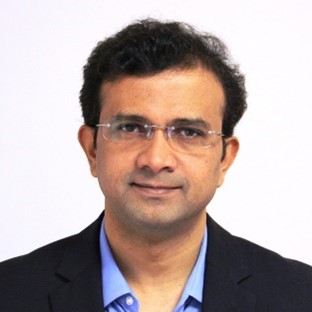}}] {Vivek Kulkarni}
He did a Master of Science in “Communications Engineering” from Technical University of Munich (TUM) in 2000 and an MBA in “Innovation and Business Creation” from TUM School of Management in 2014. In more than 20 years of industrial experience, Vivek Kulkarni worked in several industry domains such as Industrial automation, Renewable energy, Carrier and Mobile networks with focus on communication protocols, future network architecture, MPLS, GMPLS, carrier-grade Ethernet and Profinet. Since 2013, his main focus is on Software Defined Networking, Industry 4.0 and Next-Generation IoT. Currently he is the Project Coordinator of the EU H2020 project IntellIoT (Next-Generation IoT project) and in the past he was coordinator of EU-project SEMIoTICS (IoT project) as well as the EU H2020 project VirtuWind, which was one of the prestigious vertical domain projects in the 5G PPP phase-1 projects.
\end{IEEEbiography}

\begin{IEEEbiography}[{\includegraphics[width=0.9in,height=1.55in,clip,keepaspectratio]{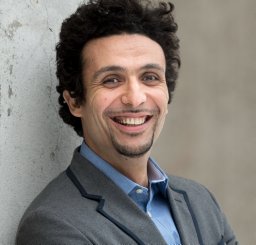}}] {Dr Mehdi Bennis}
is a full (tenured) Professor at the Centre for Wireless Communications, University of Oulu, Finland, Academy of Finland Research Fellow and head of the intelligent connectivity and networks/systems group (ICON). His main research interests are in radio resource management, heterogeneous networks, game theory and distributed machine learning in 5G networks and beyond. He has published more than 200 research papers in international conferences, journals and book chapters. He has been the recipient of several prestigious awards including the 2015 Fred W. Ellersick Prize from the IEEE Communications Society, the 2016 Best Tutorial Prize from the IEEE Communications Society, the 2017 EURASIP Best paper Award for the Journal of Wireless Communications and Networks, the all-University of Oulu award for research, the 2019 IEEE ComSoc Radio Communications Committee Early Achievement Award and the 2020 Clarviate Highly Cited Researcher by the Web of Science. Dr Bennis is an editor of IEEE TCOM and Specialty Chief Editor for Data Science for Communications in the Frontiers in Communications and Networks journal. Dr Bennis is an IEEE Fellow.
\end{IEEEbiography}

\begin{IEEEbiography}[{\includegraphics[width=0.9in,height=1.55in,clip,keepaspectratio]{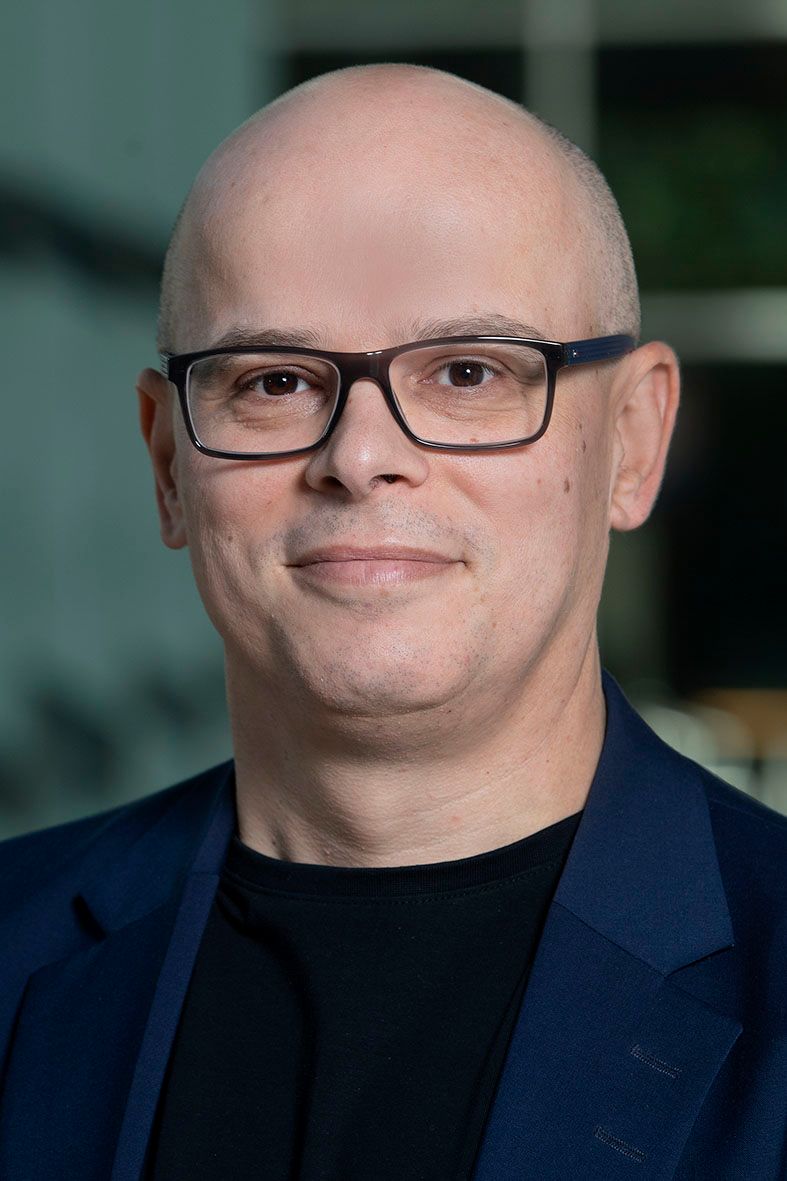}}] {Petar Popovski}
(Fellow, IEEE) is a Professor at Aalborg University, where he heads the section on Connectivity and a Visiting Excellence Chair at the University of Bremen. He received his Dipl.-Ing and M. Sc. degrees in communication engineering from the University of Sts. Cyril and Methodius in Skopje and the Ph.D. degree from Aalborg University in 2005. He is a Fellow of the IEEE. He received an ERC Consolidator Grant (2015), the Danish Elite Researcher award (2016), IEEE Fred W. Ellersick prize (2016), IEEE Stephen O. Rice prize (2018), Technical Achievement Award from the IEEE Technical Committee on Smart Grid Communications (2019), the Danish Telecommunication Prize (2020) and Villum Investigator Grant (2021). He is a Member at Large at the Board of Governors in IEEE Communication Society, Vice-Chair of the IEEE Communication Theory Technical Committee and IEEE TRANSACTIONS ON GREEN COMMUNICATIONS AND NETWORKING. He is currently an Area Editor of the IEEE TRANSACTIONS ON WIRELESS COMMUNICATIONS and, from 2022, an Editor-in-Chief of IEEEE JOURNAL ON SELECTED AREAS IN COMMUNICATIONS. Prof. Popovski was the General Chair for IEEE SmartGridComm 2018 and IEEE Communication Theory Workshop 2019. His research interests are in the area of wireless communication and communication theory. He authored the book ``Wireless Connectivity: An Intuitive and Fundamental Guide'', published by Wiley in 2020.
\end{IEEEbiography}

\end{document}

%% file: usepackage.tex
\usepackage{times}
\usepackage{amsmath,epsfig}
\usepackage{amssymb}
\usepackage{amsfonts} 
\usepackage{array,dcolumn}
\usepackage[ruled,vlined,linesnumbered]{algorithm2e}
\usepackage{subcaption}
\usepackage{psfrag}
\usepackage{amsmath}
\usepackage{amsfonts}
\usepackage{amssymb}
\usepackage{balance}
\usepackage{accents}
\usepackage{booktabs, adjustbox}
\usepackage{cite}
\usepackage{ragged2e}
\usepackage[all]{xy}
\usepackage{dsfont}
\usepackage{hyperref}
\usepackage{url}
\usepackage{xcolor}
\usepackage[nolist]{acronym}
\usepackage{graphicx} 
\usepackage{float}
\usepackage{threeparttable}
\usepackage{cancel}
\usepackage[official]{eurosym}
\usepackage{subcaption}
\usepackage{fancyhdr}
\usepackage{comment}
\usepackage{caption}
\usepackage{subcaption}
\usepackage{chemformula}
\usepackage{setspace}
\usepackage[T1]{fontenc} 
\usepackage{amsmath}
\usepackage[cmintegrals]{newtxmath}
\usepackage{bm} 
\usepackage[colorinlistoftodos,prependcaption,textsize=tiny]{todonotes}
\usepackage{amsmath,amsfonts}
\usepackage{algorithmic}
\usepackage{array}
\usepackage{textcomp}
\usepackage{stfloats}
\usepackage{url}
\usepackage{verbatim}
\usepackage{graphicx}
\def\BibTeX{{\rm B\kern-.05em{\sc i\kern-.025em b}\kern-.08em
    T\kern-.1667em\lower.7ex\hbox{E}\kern-.125emX}}
\usepackage{balance}

\newcommand{\dd}{DD}

\def\({\left(}
\def\){\right)}

\def\[{\left[}
\def\]{\right]}

\newcommand{\rv}[1]{{\textcolor{black}{#1}}}

\usepackage{glossaries}

\hypersetup{
     colorlinks = true,
     linkcolor = black,
     anchorcolor = black,
     citecolor = black,
     filecolor = blue,
     urlcolor = black
     }


%% file: 1-introduction.tex
\section{Introduction}
\subsection{Towards the edge}
During the last decade, the need for connecting billions of \gls{iot} devices has driven a significant part of the design of computing and communication networks. The number of use cases is countless, ranging from smart home to smart city, industrial automation or smart farming. Many of the applications involve huge amounts of data, and the need for fast, trustworthy and reliable processing of this data is oftentimes infeasible with a cloud-centric paradigm~\cite{Hou2016cloud, Alnoman2019edge}. Moreover, typical hierarchical setups of IoT cloud platforms hinder use cases with dynamically changing context due to lacking self-awareness of the individual subsystems and the overall system they usher. Alternatively, the architectures are evolving towards edge solutions that place compute, networking, and storage in close proximity to the devices. At the same time, the introduction of machine-driven intelligence has led to the term \textit{edge intelligence}, referring to the design of distributed \gls{iot} systems with latency-sensitive learning capabilities~\cite{Deng2020edgeintelligence}. 

Although the edge-centric approach solves the fundamental limitations in terms of latency and dynamism, it also induces new challenges to the edge system: (1) the system has to deal with complex IoT applications which include functions for sensing, acting, reasoning and control, to be collaboratively run in heterogeneous devices, such as edge computers and resource-constrained devices, and generating data from a huge number of data sources; (2) trustworthiness is a big concern for edge and \gls{iot} systems where devices communicate with other devices belonging to potentially many different parties, without any pre-established trust relationship among them; (3) all those functionalities are increasingly based on a resource-limited wireless infrastructure that introduces latency and packet losses in dynamically changing channels. 

Another huge concern for the exponential growth of \gls{iot} is its scalability and contribution to the carbon footprint. On the one hand, \gls{iot} is key in deploying a huge amount of applications that will reduce the emissions of numerous sectors and industries (e.g., smart farming or energy)~\cite{Hilty2016ICTsustainability}. On the other hand, although many of these devices are low-power, the total energy consumption of the infrastructure that support such systems does have a contribution to the digital carbon footprint and cannot be overlooked~\cite{shiftproject2019, Bol2015sustainableIoT}.  

\subsection{Intelligent IoT environments}

We coin the term \emph{\gls{iiote}} to refer to autonomous \gls{iot} applications endowed with intelligence based on an efficient and reliable IoT/edge- (computation) and network- (communication) infrastructure that dynamically adapts to changes in the environment and with built-in and assured trust. Besides the wireless (and wired) networking to interconnect all \gls{iot} devices and infrastructure,
there are other three key (and power-hungry) technologies that enable \gls{iiote}. 
The first one is \gls{ml} and \gls{ai}, and therefore we talk about \emph{intelligent} \gls{iot} environments, comprising heterogeneous devices that can collaboratively execute autonomous \gls{iot} applications. Given the distributed nature of the system, distributed \gls{ml}/\gls{ai} solutions are  better suited for multi-node (multi-agent) learning.  
\emph{Edge computing} is another defining technology that provides the computation side of the infrastructure and allocates computing resources for complex \gls{iot} applications that need to be distributed over multiple, connected \gls{iot} devices (e.g., machines and \gls{agv}s). The third pillar is the \gls{dlt}: rather than traditional security mechanisms, \gls{dlt} has been identified as the most flexible solution for trustworthiness in a fully decentralized and heterogeneous scenario. 
Combined with smart contracts, it is possible for the system to autonomously control the transactions from parties without the need for human intervention. All these ingredients are necessary for a fully functional \gls{iiote}, but they have inevitably a significant contribution to the total energy footprint. Our goal is to understand the role of each technology in the performance and energy consumption of an \gls{iiote}.  

\subsection{Example: A manufacturing plant}
\begin{figure}[!t]
    \centering
    \includegraphics[width=0.99\linewidth]{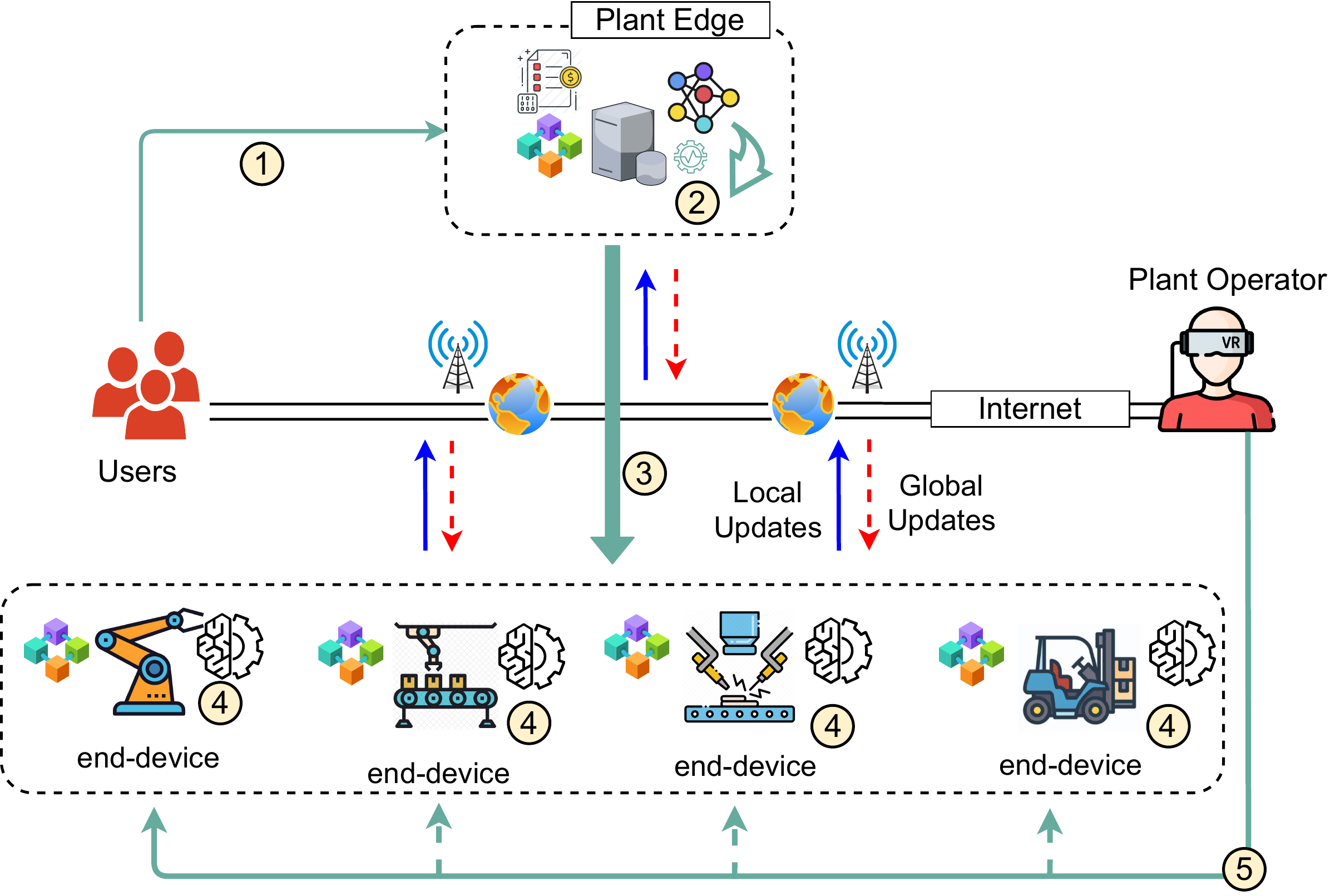}
    \caption{\gls{iiote} in a manufacturing plant.}
    \label{fig:manufacturing}
\end{figure}

A representative use case for \gls{iiote} is a manufacturing plant like shown in Fig. \ref{fig:manufacturing}, with autonomous collaboration between industrial robot arms, machinery and \gls{agv}s. 
This relies on real-time data analysis and adaptability and intelligence in the manufacturing process, which is only feasible with the edge paradigm. 
The wireless infrastructure interconnects all the machines and robots to the edge network and enables reliable and safe operation. In the figure, the following scene is depicted: a customer (the end-user) of a shared manufacturing plant orders a product by specifying a manufacturing goal (step 1). In step (2), the needed machine orchestration and associated process plan is determined to manufacture the desired product taking into account the available computation and communication resources. The event-based process planner at the edge node is responsible for observing the manufacturing process and reacting when the health state of a concerned machine changes. For example, by re-scheduling a given task in a non-responding machine. In step (3), 
the manufacturing process data is sent to the involved machines, which can include, e.g., mobile robots or an \gls{agv} to transport the work-pieces between production points, robotic arms, laser engravers, assembly stations, etc. Let us assume that the task requires a robot to pick up a work-piece and place it in different machines for its processing. As these machines may be operated by the plant owner or a third-party operator, contractual arrangements need to be set up, for which a distributed ledger is used. The ledger registers the details of each task for future accountability. 
In step (4), the local \gls{ai} on board of the different end devices comes into play. For example, in the case of the robot as an end device, its AI decides how to pick up a work-piece and place it in the next machine. 
In case the local AI of the robot cannot complete its task (e.g., because it has not been trained for a similar situation yet), a human takes over remote control (this can be e.g. a plant operator). After the human intervention, the local AI can be re-trained based on the data captured from the human input. 
This scene captures the role and interaction of the three technologies mentioned above: edge computing, ML/AI and DLTs/smart contracts. Similar examples can be defined in other domains, such as agriculture (e.g., autonomously interacting harvesting machines), healthcare (e.g., remote patient monitoring and interventions) and energy (e.g., wind plant monitoring and maintenance). 

\subsection{Contributions and outline}
In this paper, we analyze the key technologies for the next generation of \gls{iot} systems, and the tradeoffs between performance and energy consumption. We notice that characterizing the energy efficiency of these complex systems is a daunting task. The conventional approach has been to characterize every single device or link. Nevertheless, the energy expenditure of an \gls{iot} device will strongly depend on the context in which it is put, in terms of, e.g., goal of the communication or traffic behaviour. 
Therefore, we go beyond the conventional single-device approach and use the \gls{iiote} as the basic building block in the energy budget. Contrary to the single device, the \gls{iiote} is able to capture the complex interactions among devices for each of the technologies. The total energy footprint is not just a simple sum of an average per-link or per-transaction consumption of an isolated device, and scaling the number of \gls{iiote} to a large number of instances will give a more accurate picture of the overall energy consumption.

The rest of the paper is organized as follows. In Section~\ref{sec:background} we provide the state-of-the-art of the enabling technologies. Section~\ref{sec:enabling_tech} analyzes the performance and energy consumption of each enabling technology, and Section~\ref{sec:discussion} provides the vision for integrating the enabling technologies in energy-efficient \gls{iiote}. Concluding remarks and a roadmap to address the open research challenges are given in Section~\ref{sec:conclusions}.


%% file: 2-background.tex
\section{Background and Related Work}\label{sec:background}

\subsection{Edge wireless communications}

Edge computing enables the processing of the received data closer to the sensor that generated them. This means a full re-design of the communication infrastructure that must implement additional functionality at the cellular base stations or other edge nodes. The design and performance of communication networks for edge computing has been widely studied in the last years, and an overview can be found in \cite{yu2017survey} and \cite{Hassan2018edge}. One example is the term Mobile Edge Computing (MEC), adopted in 5G to refer to the deployment of cloud servers in the base stations to enable low
latency, proximity, high bandwidth, real time radio network
information and location awareness. Specifically, the concept was defined in late 2014 by the European Telecommunications Standards
Institute (ETSI): \emph{As a
complement of the C-RAN architecture, MEC aims to unite
the telecommunication and IT cloud services to provide the
cloud-computing capabilities within radio access networks
in the close vicinity of mobile users}~\cite{etsi_mec}. One of the areas of more research has been the network virtualization and slicing with the MEC paradigm~\cite{Taleb2017}. In the Radio Access Network, several authors have looked at the potential of edge computing to support \gls{urllc} \cite{Elbamby2018edge,Elbamby2019wireless,Hu2018delay}. Another research area is the use of machine learning, particularly deep learning techniques, to unleash the full potential of \gls{iot} edge computing and enable a wider range of application scenarios~\cite{Wang2020convergence,li2018learning}. However, most previous works address the communications separately. Even though several papers address the joint communication and computation resource management~\cite{Mao2017edgecomm}, they represent only the first step towards a holistic design of \gls{iiote} and its defining technologies, as well as the integration with the communication infrastructure.  

To optimize the energy efficiency of \gls{iiote}, it is interesting to choose a communication technology that ensures low power consumption and massive
connections of devices. In this regard, 3GPP introduced \gls{nbiot}, a cellular technology to utilize limited licensed spectrum of existing mobile networks to handle a limited amount of bi-directional \gls{iot} traffic. Although it uses LTE bands or guard-bands, it is usually classified as a 5G technology. It can achieve up to 250 kbps peak data rate over 180 kHz bandwidth on a LTE band or guard-band\cite{kanj2020nbiot} \cite{Wang2017}. 


Compared to other low-power technologies, \gls{nbiot} is interesting for \gls{iot} application\textcolor{blue}{s} with more frequent communications. This is the case for the ones considered in \gls{iiote}, where the intelligent end devices share the updated models frequently and must record new transactions in the ledger. At the same time, \gls{nbiot} keeps the advantages of \gls{lpwa} technologies: low power consumption and simplicity. Throughout the rest of the paper, we use \gls{nbiot} as a representative wireless technology for our analyses of \gls{iiote}. Other wireless technologies will follow similar access procedures and energy-performance trade-offs. 

\begin{figure}
    \centering
    \includegraphics[width=0.7\linewidth]{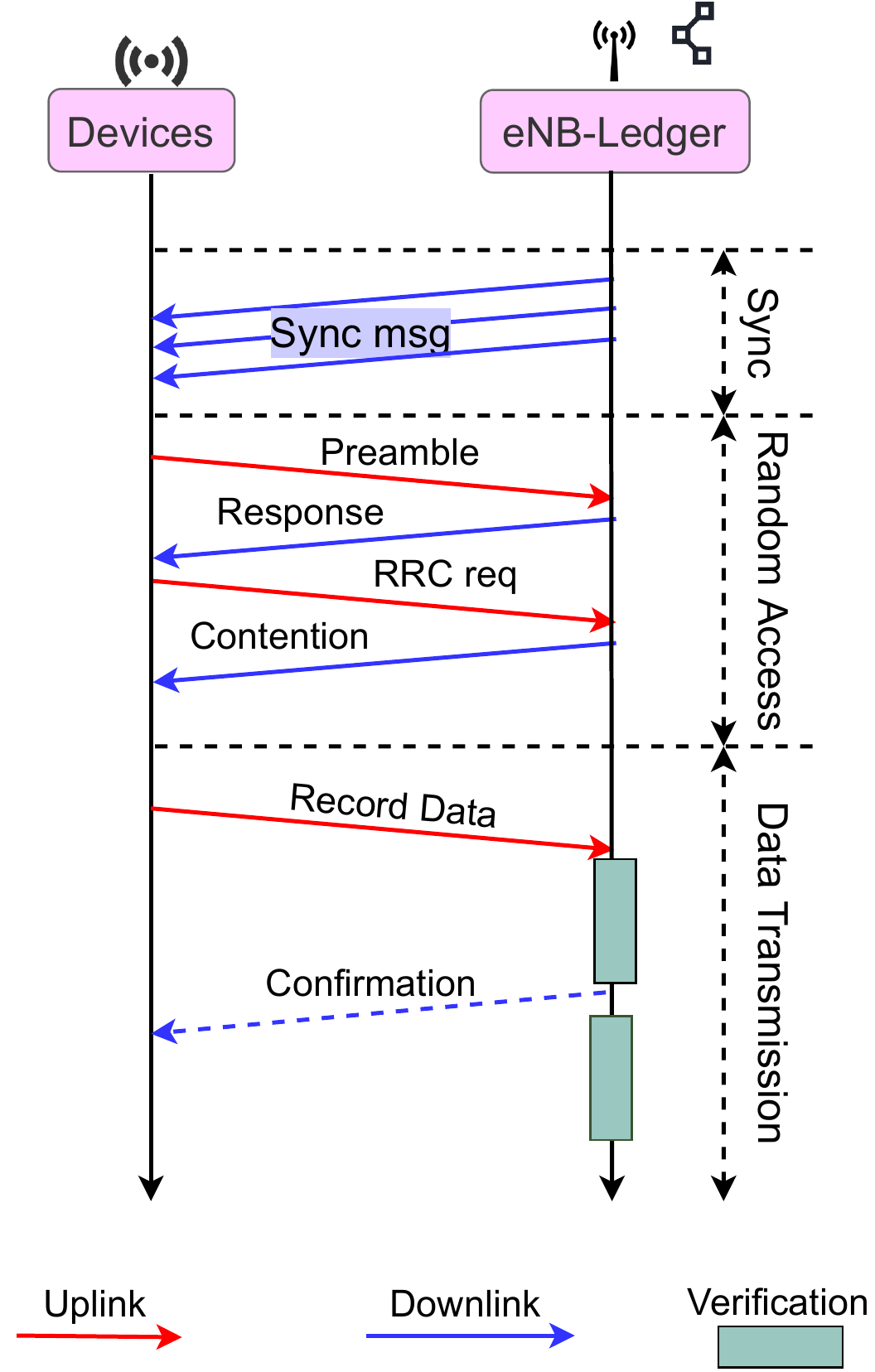}
    \caption{Random Access procedure in \gls{nbiot}}
    \label{fig:workflow}
\end{figure}

For an analysis of the energy consumption and battery lifetime of \gls{nbiot} under different configurations we refer the reader to~\cite{energymodel}. 
 A key point for this analysis is the study of the communication exchange during the access procedure: The devices that attempt to communicate through a base station must first complete a \gls{ra} procedure to transit from \gls{rrc} idle mode to \gls{rrc} connected mode. Only in \gls{rrc} connected mode data can be transmitted in the uplink through the \gls{pusch} or in the downlink through the \gls{pdsch}.  
The standard 3GPP \gls{ra} procedure consists of four message exchanges: preamble (\emph{Msg1}), uplink grant (\emph{Msg2}), connection request (\emph{Msg3}), and contention resolution (\emph{Msg4}) (see Figure~\ref{fig:workflow} where the example of recording some data, e.g., a DLT transaction is depicted). Out of these, \emph{Msg3} and \emph{Msg4} are scheduled transmissions where no contention takes place. 

The \gls{nbiot} preamble are orthogonal resources transmitted in the \gls{nprach} and used to perform the \gls{ra} request (\emph{Msg1}). A preamble is defined by a unique single-tone and pseudo-random hopping sequence. 
The \gls{nprach} is scheduled to occur periodically in specific subframes; these are reserved for the \gls{ra} requests and are commonly known as \glspl{rao}. To initiate the \gls{ra} procedure, the devices select the initial subcarrier randomly, generate the hopping sequence, and transmit it at the next available \gls{rao}. The orthogonality of preambles implies that multiple devices can access the base station in the same \gls{rao} if they select different preambles. Next, the grants are transmitted to the devices through the \gls{npdcch} within a predefined period known as the  \emph{\gls{ra} response window}. However, the number of preambles is finite and collisions can happen. In case of collision, each collided device may retransmit a preamble after a randomly selected backoff time. 

The specification provides sufficient flexibility in the configuration of the \gls{ra} process, which makes it feasible to adjust the protocol and find the right balance between reliability, latency, and energy consumption for a given application. 
Specifically, the network configures the preamble format and the maximum number of preamble transmission depending on the cell size, and this has an impact on the preamble and the total duration~\cite{3GPPTS36.331}. 
Increasing the number of preamble transmissions reduces the erasure probability, but at the cost of higher energy consumption and larger latency. The same energy-reliability-latency tradeoff applies to other messages, including the \gls{ra} response.
Moreover, scheduling the \gls{nprach} and \gls{npdcch} consumes resources that would otherwise be used for data transmission. 
Therefore, each implementation must find an adequate balance between the amount of resources dedicated to \gls{nprach}, \gls{npdcch}, \gls{pusch}, and \gls{pdsch}\footnote{It is worth mentioning that 5G has not defined a \gls{ra} procedure yet but it is expected that, when this happens, it will be heavily based on the described procedure for LTE and the energy consumption/latency tradeoff will follow similar principles. }.

\subsection{Distributed Learning over Wireless Networks}
\label{sec:background-DL}
Implementing intelligent \gls{iot} systems with distributed \gls{ml}/\gls{ai} over wireless networks (e.g., \gls{nbiot}) needs to consider the impact of the communication network (latency and reliability under communication overhead and channel dynamics) and on-device constraints (access to data, energy, memory, compute, and privacy, etc.).
Obtaining high-quality  trained models without sharing raw data is of utmost importance, and redounds to the trustworthiness of the system. 
In this view, \gls{fl} has received a groundswell interest in both academia and industry, whose underlying principle is to train a \gls{ml} model by exchanging model parameters (e.g., \gls{nn} weights and/or gradients) among edge devices under the orchestration of a federation server and without revealing raw data \cite{pap:jakub16}.
Therein, devices periodically upload their model parameters after their local training to a parameter server, which in return does model averaging and broadcasting the resultant global model to all devices.
\gls{fl} has been proposed by Google for its predictive keyboards \cite{yang2018applied} and later on adopted in different use cases in the areas of intelligent transportation, healthcare and industrial automation, and many others \cite{samarakoon2019distributed,kairouz2019advances}. 
While \gls{fl} is designed for training over homogeneous agents with a common objective,  recent studies have extended the focus towards personalization (i.e., multi-task learning) \cite{jnl:smith17}, training over dynamic topologies \cite{lalitha2019peer} and robustness guarantees \cite{ARM18,sattler2019robust}.
In terms of improving data privacy against malicious attackers, various privacy-preserving methods including injecting fine-tuned noise into model parameters via a differential privacy mechanism \cite{9170559,zhao2020local,lyu2020threats,yang2020local} and mixing model parameters over the air via analog transmissions \cite{koda2020differentially,elgabli2020harnessing} have been recently investigated.
Despite of the advancements in \gls{fl} design, one main drawback in the design of \gls{fl} is that its communication overhead is proportional to the number of model parameters calling for the design of communication-efficient \gls{fl}.
In an edge setup with limited resources in communication and computation, this introduces training stragglers degrading the overall training performance. 
In this view, client scheduling \cite{ yang2019scheduling,wadu2020federated,bonawitz2019towards} and computation offloading \cite{amiri2019computation,Elbamby:2019,polese2020machine} with the focus on guaranteeing target training/inference accuracy have been identified as a promising research direction.
With client scheduling, the number of communication links are reduced (known as \emph{link sparcification}) and thus, the communication bandwidth and energy consumption of distributed learning can be significantly decreased.
Additional temporal link sparsity can be introduced by enforcing model sharing policies that account model changes and/or importance within consecutive training iterations such as the \gls{lag} method \cite{chen2018lag}.
Sparsity can be further exploited by adopting sparse network topologies, which rely on communications within a limited neighborhood in the absence of a central coordinator/helper. 
While such sparsification improves energy and communication efficiencies, it could yield higher learning convergence speed as well as lower training and inference accuracy, in which sparsity needs to be optimized in terms of the trade-off between communication cost and convergence speed.
In this view, several sparse-topology-based distributed learning methods including decentralized \gls{gd}, dual averaging \cite{duchi2011dual}, learning over graphs \cite{alshammari2021, wang2019matcha} and GADMM algorithms \cite{elgabli2020, issaid2020} have been investigated.
%

%

\subsection{Optimizing \gls{iot} Application Deployments in \gls{iot} Environments}
\label{sec:optimal-allocation}

\gls{iot} applications typically consist of multiple \emph{components}. For instance, an \gls{iot} application could comprise components for secure data acquisition (e.g., based on Blockchain), data pre-processing, feeding the data into a neural network (or even through multiple ones) before it acts upon the outcome of the \gls{ml} inference, etc. In many cases, such composed \gls{iot} applications need to be distributed over multiple, connected intelligent \gls{iot} devices. An important aspect is then to optimize this allocation of application components to devices. The result of the allocation is an assignment of components to devices, that fulfills the constraints, and optimizes the performance of the system in some metric. This metric could, for example, maximize the responsiveness of the application or minimize the overall energy consumption, where the latter is reasonable in battery-run wireless systems. 
An overview of existing allocation approaches is given in~\cite{lakshmanan_placement_2008}.

Previous work \cite{Haubenwaller2015} used Constraint Programming to describe an approach for the efficient distribution of actors to \gls{iot} devices.
The approach resembles the \textcolor{black}{Quadratic Assignment Problem (QAP)} and is NP-hard, resulting in long computation times when scaling up.
Samie et al.~\cite{Samie2016} present another Constraint Programming-based approach that takes into account the bandwidth limitations and minimizing energy consumption of \gls{iot} nodes.
The system optimizes computation offloading from an \gls{iot} node to a gateway, however, it does not consider composed computations that can be distributed to multiple devices.

A Game Theory-based approach is presented in~\cite{Sardellitti2015} that aims at the joint optimization of radio and computational resources of mobile devices.
However, the system local optimum for multiple users only aims at deciding whether to fully offload a computation or to fully process it on device.

Based on Non-linear Integer Programming, Sahni et al.~\cite{Sahni2017} present their Edge Mesh algorithm for task allocation that optimizes overall energy consumption and considers data distribution, task dependency, embedded device constraints, and device heterogeneity.
However, only basic evaluation and experimentation are done, without performance comparison.
Based on Integer Linear Programming (ILP), Mohan \& Kangasharju~\cite{Mohan2016} propose a task assignment solver that first minimizes the processing cost and secondly optimizes the network cost, which stems from the assumption that Edge resources may not be highly processing-capable.
An intermediary step reduces the sub-problem space by combining tasks and jobs with the same associated costs. This reduces the overall processing costs.

Cardellini et al.~\cite{cardellini16optimal_operator_placement} describe a comprehensive ILP-based framework for optimally placing operators of distributed stream processing applications, while being flexible enough to be adjusted to other application contexts. Different optimization goals are considered, e.g., application response time and availability. They propose their solution as a unified general formulation of the optimal placement problem and provide an appropriate theoretical foundation. The framework is flexible so that it can be extended by adding further constraints or shifted to other optimization targets.
Finally, our previous work \cite{seeger2020optimally} has leveraged Cardellini's framework and has extended it by incorporating further constraints for the optimization goal, namely the overall energy usage of the application.  

\subsection{Distributed Ledger Technologies over Wireless Networks}

In recent years, \gls{dlt} has been the focus of large research efforts spanning several application domains. Starting with the adoption of Bitcoin and Blockchain, \gls{dlt} has received a lot of attention in the realm of \gls{iot}, as the technology promises to help address some of the \gls{iot} security and scalability challenges \cite{fernandez2018review}.
%
%
For instance, in \gls{iot} deployments, the recorded data are either centralized or spread out across different heterogeneous parties. These data can be both public or private, which makes it difficult to validate their origin and consistency. In addition, querying and performing operations on the data becomes a challenge due to the incompatibility between different Application Programming Interfaces (APIs). For instance, Non-Governmental Organizations (NGOs), Public and Private sectors, and industrial companies may use different data types and databases, which leads to difficulties when sharing the data \cite{9141220}. 
%
%
A \gls{dlt} system offers a tamper-proof ledger that is distributed on a collection of communicating nodes, all sharing the same initial block of information, the genesis block \cite{nakamoto2019bitcoin}. In order to publish data to the ledger, a node includes data formatted in transactions in a block with a pointer to its previous block, which creates a chain of blocks, the so called Blockchain.
%

A smart contract \cite{wood2014ethereum} is a distributed app that lives in the Blockchain. This app is, in essence, a programming language class with fields and methods, and they are executed in a transparent manner on all nodes participating in a Blockchain \cite{christidis2016blockchains}. Smart contracts are the main blockchain-powered mechanism that is likely to gain a wide acceptance in \gls{iot}, where they can encode transaction logic and policies, which includes the requirements and obligations of parties requesting access, the \gls{iot} resource/service provider, as well as data trading over wireless \gls{iot} networks \cite{9324804}. 
With the aforementioned characteristics, the advantages of the integration of \gls{dlt}s into wireless \gls{iot} networks consist of: i) guarantee of immutability and transparency for recorded \gls{iot} data; ii) removal of the need for third parties; iii) development of a transparent system for heterogeneous \gls{iot} networks to prevent tampering and injection of fake data from the stakeholders.


\gls{dlt}s have been applied in various \gls{iot} areas such as healthcare \cite{esposito2018blockchain, griggs2018healthcare}, supply chain \cite{saberi2019blockchain}, smart manufacturing \cite{zhang2019blockchain}, and vehicular networks \cite{yang2018blockchain}. In the smart manufacturing area, the work described in \cite{zhang2019blockchain} investigates \gls{dlt}-based security and trust mechanisms and elaborates a particular application of \gls{dlt}s for quality assurance, which is one of the strategic priorities of smart manufacturing. Data generated in a smart manufacturing process can be leveraged to retrieve material provenance, facilitate equipment management, increase transaction efficiency, and create a flexible pricing mechanism. 
%

%
%
One of the challenges of implementing \gls{dlt} in \gls{iot} and edge computing is the limited computation and communication capabilities of some of the nodes. In this regard, the authors in \cite{danzi2020communication, 9324804} worked on the communication aspects of integrating \gls{dlt}s with \gls{iot} systems. The authors studied the trade-off between the wireless communication and the trustworthiness with two wireless technologies, LoRa and \gls{nbiot}. 


%% file: 4-enabling-technologies.tex
\section{Enabling technologies for \gls{iiote}} \label{sec:enabling_tech}

This section elaborates on the three enabling technologies for \gls{iiote}: distributed learning, distributed computing, and distributed ledgers.

\subsection{Energy-efficient Distributed Learning over Wireless Networks}
\label{sec:enabling_tech-DL}
As shown in Figure~\ref{fig:manufacturing}, each end device in the \gls{iiote} has local \gls{ai} and the whole system relies on \gls{fl}. We present learning frameworks that are suitable for \gls{iiote} leveraging two techniques: (1) spatial and temporal sparsity; and (2) quantization.

\subsubsection{Dynamic GADMM} \label{sec:D-GADMM}
Standard \gls{fl} requires a central entity, which plays the role of a parameter server (PS). At every iteration, all nodes need to communicate with the PS, which may not be an energy-efficient solution especially for a large distributed network of agents/workers, as in the manufacturing use case. Furthermore, a PS-based approach is vulnerable to a single point of attack or failure. 
To overcome this problem and ensure a more energy-efficient solution, we propose a variant of the standard \gls{admm}~\cite{boyd2011distributed} method that decomposes the problem into a set of subproblems that are  solved in parallel, referred to as \gls{gadmm}\cite{elgabli2020}.
\Gls{gadmm} extends the standard \gls{admm} to decentralized topology and enables communication and energy-efficient distributed learning by leveraging spatial sparsity, i.e. enforcing each worker to communicate with at most two neighbouring workers. In \gls{gadmm}, the standard learning problem (\textbf{P1}) is re-formulated as the following learning problem (\textbf{P2}):
%
%
\begin{align}
    \label{P1}
   (\textbf{P1}) & \textstyle \quad \min_{\{\boldsymbol{\theta}_n\}_{n=1}^N} \sum\limits_{n=1}^N f_n(\boldsymbol{\theta}_n)
\end{align}
\begin{align}
   \label{P2}
   \begin{split}
   (\textbf{P2}) \textstyle \quad \min_{\{\boldsymbol{\theta}_n\}_{n=1}^N} \sum\limits_{n=1}^N f_n(\boldsymbol{\theta}_n) \qquad \\
    \text{s.t.~} \boldsymbol{\theta}_{n} = \boldsymbol{\theta}_{n+1}, \text{~for~} n=1,\cdots, N-1.
   \end{split}
\end{align}

To this end, \gls{gadmm} divides the set of workers into two groups {\it head} and {\it tail}. Thanks to the equality constraint of (\textbf{P2}), each worker from the head/tail group exchanges model with only two workers from the tail/head group forming a chain topology. At iteration $k+1$, giving the models of the tail workers and the dual variables at iteration $k$, all head workers update their models in parallel since they have no joint constraints. Once the head workers update their models, they transmit their updated model to their neighbours from the tail group. Then, following the same way, every tail worker updates its model. Finally, the dual variables are updated locally at each worker. Following this alternation, \gls{gadmm} allows at most $N/2$ workers to compete over the available bandwidth compared to $N$ workers for the PS-based approach. With that, \gls{gadmm} can significantly increase the bandwidth available to each worker, which reduces the energy wasted in competition for communication resources. The energy expenditure for communication is further reduced by including only two neighbouring workers. The detailed algorithm is described in~\cite{elgabli2020}. 

One drawback of \gls{gadmm} is attributed to its slow convergence compared to standard \gls{admm}. In other words, due to the sparsification of the graph, workers require more iterations for the convergence.
To alleviate this issue and  combine the fast convergence of standard \gls{admm} with the communication-efficiency of \gls{gadmm}, we have proposed \gls{dgadmm}~\cite{elgabli2020}. Not only \gls{dgadmm} improves the convergence speed of \gls{gadmm}, but it also copes with dynamic (time-variant) networks, in which the workers are moving (e.g., the \gls{agv}s in the manufacturing plant or the tractors in the agriculture use case), while inheriting the theoretical convergence guarantees of \gls{gadmm}. In a nutshell, every couple of iterations in \gls{dgadmm}, i.e. system coherence time, two things are changing: (i) workers assignment to head/tail group, which follows a predefined assignment mechanism and (ii) neighbours of each worker from the other group. \textcolor{black}{The idea at high level as is follows: the workers are given fixed IDs, and they share a pseudo-random code that is used every $\tau$ seconds, where $\tau$ is the system coherence time to generate a set of random integers with cardinality $N/2-2$. If $n$ belongs to the set, then worker $n$ is a head worker for this period. The assumption is that workers $1$ and $N$ do not change their assignment. i.e., worker $1$ is always a head and worker $N$ is always a tail. Head workers broadcast their IDs alongside a pilot signal, then tail workers compute their communication cost to all head workers, and share the cost vector with the neighboring heads. If a tail does not receive a signal from a certain head, the cost to that head is $\infty$, the same applies to heads.  Subsequently, every head locally computes the communication-efficient chain using a predefined heuristic and share it with its neighboring tails. This approach requires two communication rounds and guarantees that every head will compute the same chain. Once the chain information is calculated, each worker will share its right dual variable with its right neighbor to be used by both workers and GADMM continues for $\tau$ seconds. It is worth mentioning that we could, e.g., start with a chain $1-2-3-\cdots-N$ and move to $1-5-7-4-\cdots-N$, so only nodes $1$ and $N$ preserve their assignments.} For further details, the reader is referred to \cite{elgabli2020} where a comprehensive explanation of the steps of \gls{dgadmm} can be found.

\begin{figure}[!t]
    \centering
    \includegraphics[width=0.8\linewidth]{ 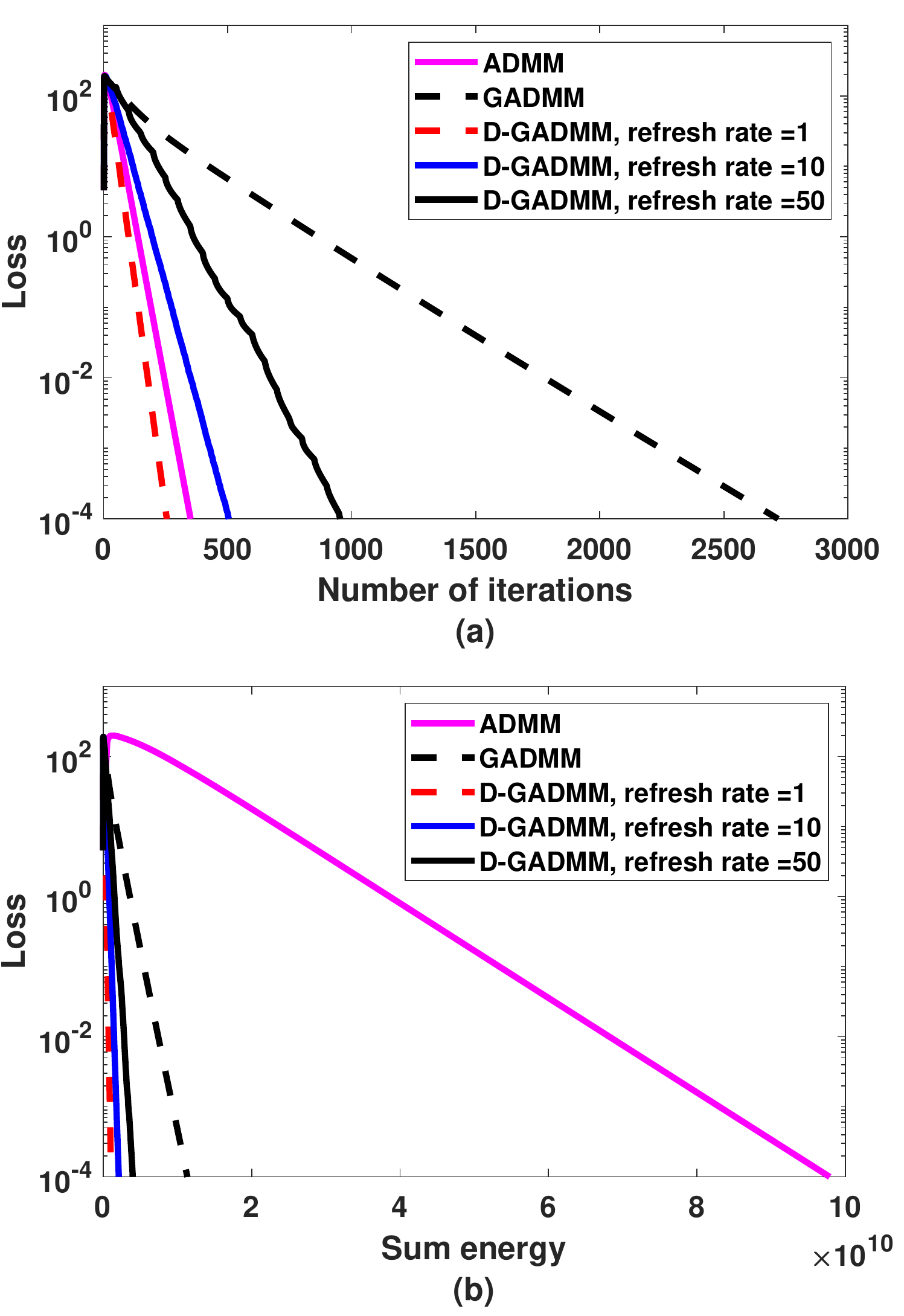}
    \caption{\Gls{dgadmm}: loss as a function of (a) number of iterations and (b) total energy consumption.}
    \label{D-GADMM}
\end{figure}

In Fig. \ref{D-GADMM}, we plot the objective error in terms of the number of iterations (left) and in terms of sum energy (right) for \gls{dgadmm} as well as \gls{gadmm} and standard \gls{admm}. As we can see from Fig. \ref{D-GADMM}, \gls{dgadmm} greatly increases the convergence speed of \gls{gadmm} and thus decreases the overall communication cost for fixed topology. As a consequence, \gls{dgadmm} achieves convergence speed comparable to the PS-based \gls{admm} while maintaining \gls{gadmm}'s low communication cost per iteration.

\subsubsection{Censored Quantized Generalized \gls{gadmm}} \label{sec:C-GGADMM}
As pointed out earlier, each worker, in the \gls{gadmm} framework, exchanges its model with  up to two neighbouring workers only, which slows down convergence. To reduce the communication overhead while generalizing to more generic network topologies, we propose the \gls{ggadmm} \cite{issaid2020}. Under this generalized framework, the workers are still divided into two groups: head and tail, with possibly different sizes. In other words, the topology is generalized from a chain topology to any bipartite graph where the number of neighbours, that each worker can communicate with can be any arbitrary number and not necessarily limited to two. By leveraging the censoring idea, i.e. temporal sparsity, we introduce the \gls{cggadmm} where each worker exchanges its model only if the difference between its current and previous models is greater than a certain threshold. To make the algorithm more communication-efficient, censoring is applied on the quantized version of the worker's model instead of the model itself to get the \gls{cqggadmm} \cite{issaid2020, elgabli2020q}. \Gls{cqggadmm} can significantly reduce the communication overhead, particularly for large model size $d$, since its payload size is $(bd + 32)$ bits compared to the payload size of $32 d$ bits for the full precision \gls{ggadmm}. Since according to the Shannon's capacity theorem, more bits consume more transmission energy for the same bandwidth, transmission duration, and noise spectral density, the communication energy of \gls{cqggadmm}, compared to the original \gls{gadmm}, is significantly reduced. Theoretically, \gls{cqggadmm} inherits the same performance and convergence guarantees of vanilla \gls{ggadmm}, provided that the censoring threshold sequence is non-increasing and non-negative. 

\begin{figure}[!t]
    \centering
    \includegraphics[width=0.8\linewidth]{ 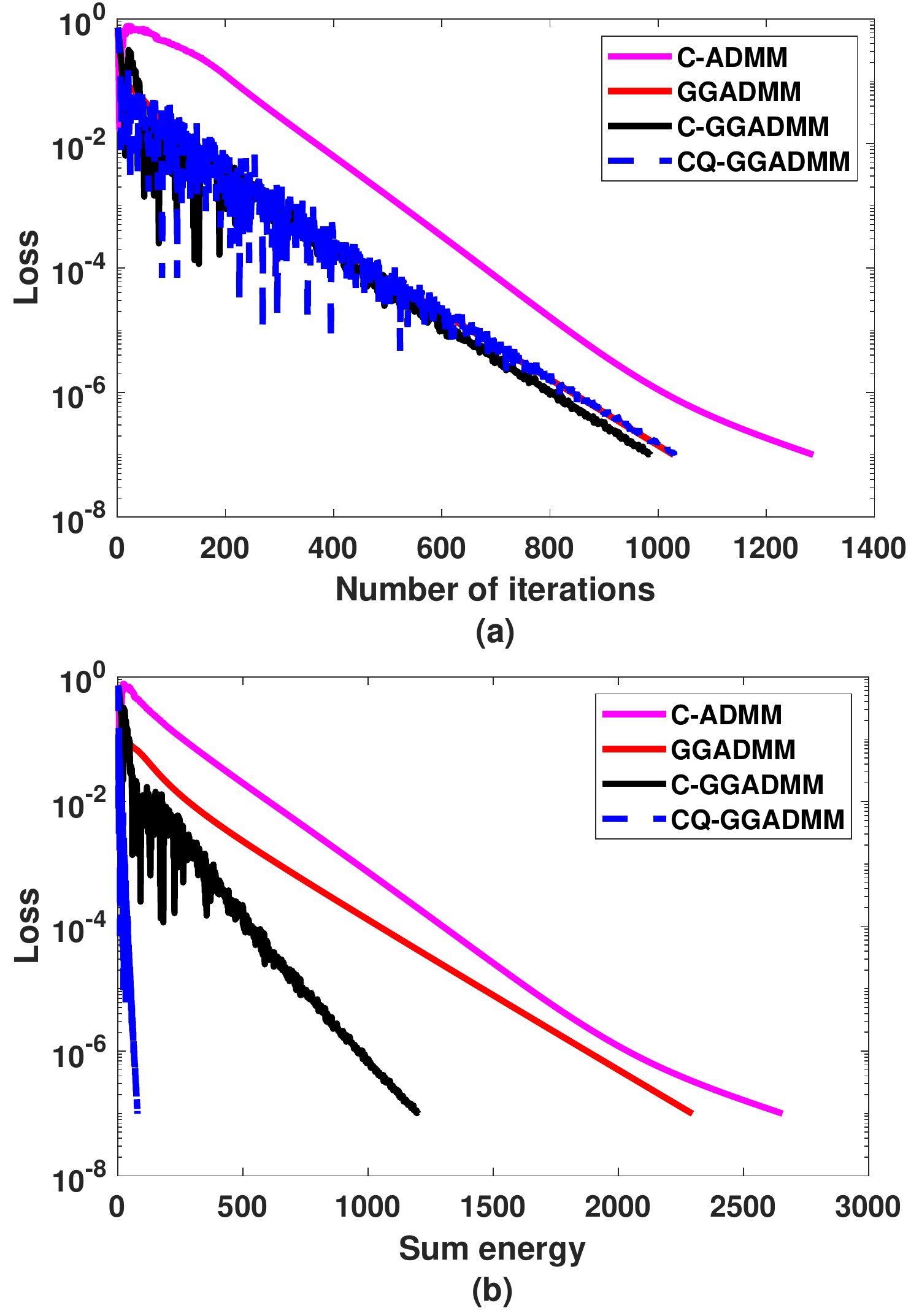}
    \caption{\Gls{cqggadmm}: loss as a function of (a) number of iterations and (b) total energy consumption.}
    \label{CQ-GGADMM}
\end{figure}

Fig. \ref{CQ-GGADMM} compares \gls{cqggadmm} with \gls{cadmm}, \gls{ggadmm}, as well as \gls{cggadmm} in terms of the loss versus the number of iterations (left) and versus the total sum energy (right) for a system of $18$ workers on a linear regression task using the Body Fat dataset \cite{Dua:2019}.
\textcolor{black}{We can observe, from Fig. \ref{CQ-GGADMM}, that \gls{cqggadmm} exhibits the lowest total communication energy, followed by \gls{cggadmm}, then \gls{ggadmm} and finally \gls{cadmm}, while having similar convergence speed to \gls{ggadmm}.} 
This observation validates the benefits of censoring the quantized version of the models before sharing, which makes the proposed algorithm (\gls{cqggadmm}) more communication and energy efficient.

Finally, it is worth mentioning that motivated by the fact that, in \gls{fl}, the parameter server is interested in the aggregated output of all workers rather than the individual output of each worker, analog over the air aggregation schemes such as \cite{amiri2019over, zhu2019broadband, sery2020analog, elgabli2020harnessing} were proposed. Such schemes were shown to achieve high scalability and significant savings in  energy consumption owing to their ability to allow non-orthogonal access to the bandwidth.

\subsection{Optimizing Energy Consumption of Wireless \gls{iot} Environments}
\label{sec:enabling_tech-Optimizing}


The next pillar in \gls{iiote} is edge computing. Specifically, we consider the problem of allocating the application components to the available end devices. As first presented in  \cite{seeger2020optimally}, we extend the Integer Linear Programming (ILP) based framework defined by Cardellini et al.~\cite{cardellini16optimal_operator_placement} (Section \ref{sec:optimal-allocation}). In \cite{seeger2020optimally} the goal was to minimize the overall energy consumption needed for executing an IoT application. The formulated ILP model is described below. The optimality can be determined with this by feeding it into a solver, such as IBM CPLEX\footnote{\url{https://www.ibm.com/analytics/cplex-optimizer}}.

We define optimality of the allocation by total energy use over one execution of an \gls{iot} application.
Energy during the application's execution is consumed in two phases: (1) \emph{device energy}, consumed by a device when executing a component; (2) and \emph{edge network energy}, consumed by the device when sending the result of the calculation over the network.
Note that ``optimal'' in this case only describes optimality in the integer model.
Given the constraints and the model, we find the optimal assignment, i.e.\ the one with minimal energy usage.


The optimal network configuration is the assignment of application components to devices that result in the lowest total consumption of energy and satisfies the constraints.
The constraints concern the requirements that an assignment must satisfy: Each component should only be allocated once and resource requirements for assigned components should not exceed the resources of the node.
This problem is a form of the quadratic assignment problem, and  thus is NP-hard.


\subsubsection{System model}
The application consists of a set of components and edges that interconnect them, modeled as a weighted undirected graph $\mathcal{G}_{\text{app}}=\left(\mathcal{V}_{\text{app}}, \mathcal{E}_{\text{app}}\right)$. Graph $\mathcal{G}_{\text{app}}$ is multi-partite, with vertex set $\mathcal{V}_{\text{app}}$ containing the application components, $|\mathcal{V}_\text{app}|=N$, and edge set $\mathcal{E}_{\text{app}}\subset \left\{t_1 t_2: t_i\in\mathcal{V}_\text{app}, i=1,2\right\}$ representing the logical connections between components $t_i$.

Analogously, the network infrastructure where the components can be evaluated is modeled with the multi-partite graph $\mathcal{G}_{\text{net}}=\left(\mathcal{V}_{\text{net}}, \mathcal{E}_{\text{net}}\right)$ with vertex set $\mathcal{V}_{\text{net}}$ containing the communicating nodes, with cardinality $|\mathcal{V}_\text{net}|=M$, and edge set $\mathcal{E}_{\text{net}}\subset \left\{t_1 t_2: t_i\in\mathcal{V}_\text{net}, n=1,2\right\}$ representing the wireless and wired links among nodes $n_i$. 
The result of the allocation is a matrix $X = \mathcal{V}_{\text{app}} \times \mathcal{V}_{\text{net}}$ where $X[t, n] = 1$ if and only if component $t$ is allocated to node $n$.
We also define $E_d$ to be the device energy and $E_n$ the network energy. $E_t$ is then the total energy, and we put the constraint $E_d + E_n \leq E_t$.
Components, nodes and links have properties that are relevant for the energy consumption of the application once allocated.
These parameters are described in Table~\ref{tab:allocation-parameters}.
$S_t$, $P_n$, $R_n$ and $C_n$ are defined as multiples of some reference node.
The resources of a node are expressed as a single scalar, but additional resource requirements can easily be introduced into the model.

\begin{table*}
  \centering
    \caption{\label{tab:allocation-parameters}Parameters of energy-aware allocation algorithm.}
  \begin{tabular}{ll}
    \toprule
    Symbol & Description\\
    \midrule
    $R_t$ & Resources required for the evaluation of component $t \in \mathcal{V}_{\text{app}}$.\\
    $O_t$ & Output of component $t \in \mathcal{V}_{\text{app}}$ for a single received input.\\
    $S_t$ & Computation time required for completing component $t \in \mathcal{V}_{\text{app}}$ once.\\
    $P_n$ & Processing power of node $n \in \mathcal{V}_{\text{net}}$.\\
    $R_n$ & Resources available on node $n \in \mathcal{V}_{\text{net}}$.\\
    $C_n$ & Energy consumption of node $n \in \mathcal{V}_{\text{net}}$ for one unit of computation.\\
    $T_l$ & Energy use for the transfer of one data packet over link $l \in E_{\text{net}}$.\\
    $D_{(n_1, n_2)}$ & Energy cost of the shortest path between $n_1$ and $n_2$.\\
    \bottomrule
  \end{tabular}
\end{table*}

\subsubsection{Problem formulation}
For calculating the network energy, we need to know whether a link between two components is assigned to a link between two nodes.
For this, we introduce a matrix $Y= \mathcal{V}_{\text{app}} \times \mathcal{V}_{\text{app}} \times \mathcal{V}_{\text{net}} \times \mathcal{V}_{\text{net}}$, where $Y[t_1, t_2, n_1, n_2] = 1$ if and only if the communication between component $t_1$ and component $t_2$ is allocated on the network link between nodes $n_1$ and $n_2$.
This corresponds to $X[t_1, n_1] = 1 \wedge X[t_2, n_2]$.
Unfortunately, this is not a linear constraint, and thus we need to linearize the formulation. For this, we follow the formulation presented in~\cite{cardellini16optimal_operator_placement} and define an ILP model as:

\begin{align}
  \forall t_1, t_2 \in \mathcal{V}_{\text{app}}: \forall n_1, n_2 \in \mathcal{V}_{\text{net}}: Y[t_1, t_2, n_1, n_2] &\le X[t_1, n_1]\label{eq:10}\\
  \forall t_1, t_2 \in \mathcal{V}_{\text{app}}: \forall n_1, n_2 \in \mathcal{V}_{\text{net}}: Y[t_1, t_2, n_1, n_2] &\le X[t_2, n_2]\label{eq:11}
\end{align}
\begin{align}
    \begin{split}
          \forall t_1, t_2 \in \mathcal{V}_{\text{app}}: \forall n_1, n_2 \in \mathcal{V}_{\text{net}}: Y[t_1, t_2, n_1, n_2] \\ \ge X[t_1, n_1] + X[t_2, n_2] - 1\label{eq:12}
    \end{split}
\end{align}
\begin{align}
  \forall t \in \mathcal{V}_{\text{app}}: \sum_{n \in \mathcal{V}_{\text{net}}} X[t,n] &= 1\label{eq:6}\\
  \forall n \in \mathcal{V}_{\text{net}}: \sum_{t \in \mathcal{V}_{\text{app}}} X[t,n] \cdot R_t &\le R_n\label{eq:7}\\
  \sum_{t \in \mathcal{V}_{\text{app}}}\sum_{n \in \mathcal{V}_{\text{net}}} C_n \cdot (S_t / P_n) \cdot X[t,n] &\le E_d\label{eq:8}\\
  \sum_{(t_1, t_2) \in \mathcal{E}_{\text{app}}} \sum_{n_1, n_2 \in \mathcal{V}_{\text{net}}} O_{n_1} \cdot P_{n_1,n_2} \cdot Y[t_1, t_2, n_1, n_2] &\le E_n\label{eq:9}\\
  E_n + E_d &\le E_t\label{eq:14}
\end{align}

\noindent where equations~(\ref{eq:10}) to~(\ref{eq:12}) describe the linearization of the network matrix $Y$.
(\ref{eq:6}) and~(\ref{eq:7}) are for ensuring that components are allocated only once and that resources are not exceeded, respectively. 
Equations~(\ref{eq:8}) and~(\ref{eq:9}) calculate network and device energy as described above.
Finally, we calculate the total energy use of the assignment by adding both energies in (\ref{eq:14}).
The objective of the optimization is the minimization of the total used energy.

\subsubsection{A Linear Heuristic for Energy-Optimized Allocation}
\label{sec:optim-energy-optim}

The presented QAP is NP-hard and thus compute intensive.
The culprit for this is the network cost calculation and the linearization of $Y$ resulting in a large number of constraints.
By removing the $Y$ matrix and the associated constraints, we create a linear problem that can be evaluated effectively by the simplex method~\cite{nelder_simplex_1965}.
The approach approximates the energy required for sending a packet of data by taking the average of a node's links.
We introduce the parameter $\hat{T}_n = \frac{1}{|\text{outgoing}(n)|} \sum_{e \in \text{outgoing}(n)} T_e$ that describes the average transmission cost of a node's links.

\begin{align}
  \sum_{t \in \mathcal{V}_{\text{app}}}\sum_{n \in \mathcal{V}_{\text{net}}} C_n \cdot (S_t / P_n) \cdot X[t,n] + O_{t} \cdot \hat{T}_n \cdot X[t,n] &\le E_t\label{eq:13}
\end{align}

The complete model reuses constraints in equations~(\ref{eq:6}) and~(\ref{eq:7}) with the constraint (\ref{eq:13}).
By transforming the QAP into a linear problem, we greatly increase the speed of finding a solution, and make the optimization feasible for on-line usage. The drawback is that by approximating the network energy the solution is no longer optimal, as it will be shown in the results. 

\subsubsection{Evaluation of Allocation Algorithm}\label{sec:evaluation-mitigator}
We implemented the model using the PuLP\footnote{\url{https://pythonhosted.org/PuLP/}} linear programming library.
The evaluation was done by generating a random network and a random application, and letting the solver find the optimal allocation.

\begin{figure}[t!]
  \centering
  \includegraphics[width=0.9\linewidth]{ 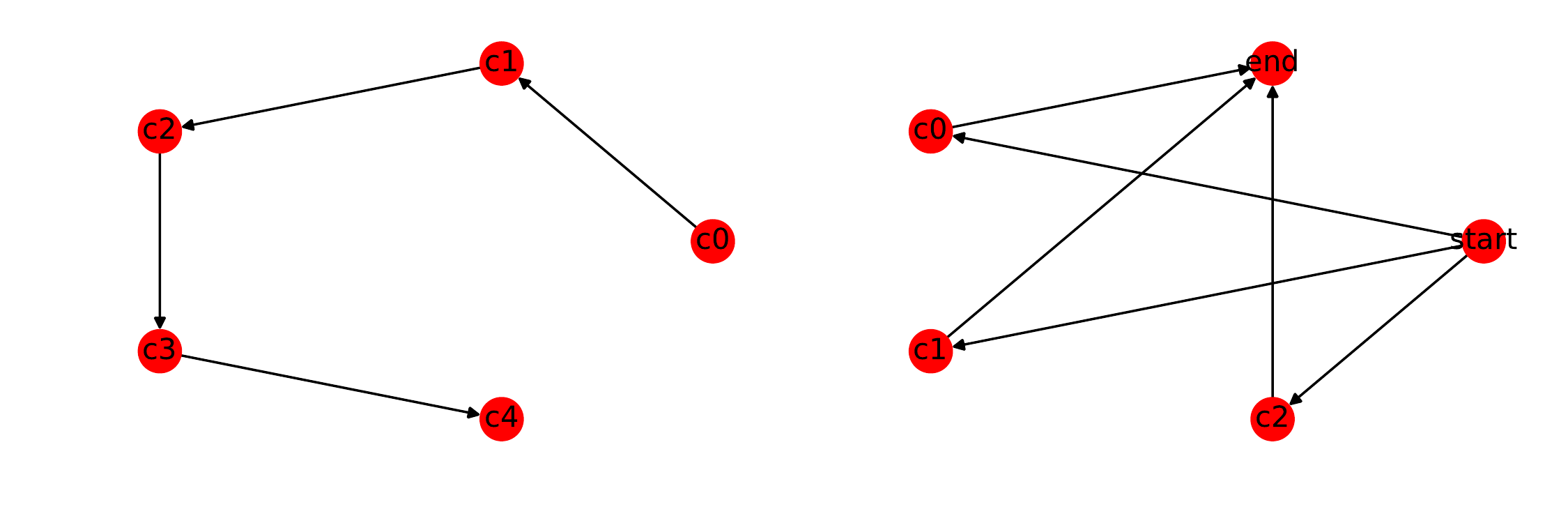}
  \caption{\label{fig:long-wide-recipes} ``Long'' (left) and ``wide'' (right) composed \gls{iot} applications.}
\end{figure}

The network configuration is generated with a variety of node configurations and capabilities, reflecting a heterogeneous computation and communication infrastructure that one could find in an industrial manufacturing plant (e.g., using Siemens range of industrial computers \cite{Siemens2019}). 
In the evaluated configuration, 60\% of the nodes were generated as wired nodes, and the remaining 40\% are wireless nodes.
Nodes are connected to each other with a certain probability.
That probability is 0.8 for wired-wired connections, 0.5 for wireless-wireless connections and 0.4 for wireless-wired connections.
Wired connections use 0.2 units of energy, while wireless connections use 0.8 units of energy,  which is similar to the power consumption of an Ethernet module \cite{ethernetModuleSpec} as compared to a WiFi module \cite{wlanPowerComparison}.
Nodes have a varying amount of memory resources uniformly distributed between a lower bound of 1 and an upper bound of 8 resource units.
Nodes also have a varying processing speed between 1 and 3 speedup, roughly comparing to the Intel processor family i3, i5, and i7.
Finally, nodes can use from 0.5 to 1.5 units of energy for a single unit of computation.

For the application, two classes with a certain number of components are generated, a ``wide'' and a ``long'' application.
In a ``wide'' application, two components are designated the ``start'' and ``end'' components, and every other component needs input from the start node and sends output to the end node.
In a long application, components are linked serially.
Figure~\ref{fig:long-wide-recipes} shows two example applications.
This method for generating recips is similar to\cite{cardellini16optimal_operator_placement}.
Each application component has resource requirements randomly distributed between 1 and 8, an output factor randomly distributed between 0.5 and 1.5, and a computation size of 1 or 2.

\begin{figure}
  \centering
  
  \begin{subfigure}{.45\textwidth}
    \includegraphics[width=0.99\textwidth]{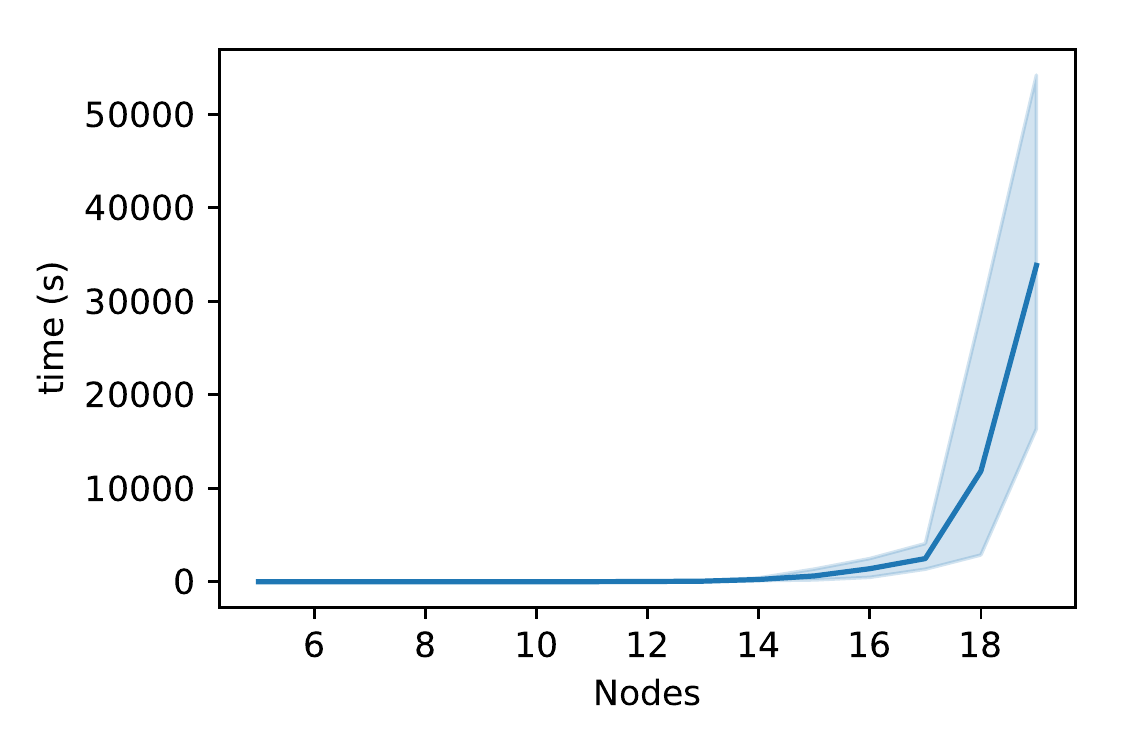}
    \caption{\label{fig:runt-optim-alloc-nodes} CPU time of the optimal allocation algorithm vs.\ the number of nodes.
      Each experiment with n nodes was measured 5 times with 3 to n-1 components.}
  \end{subfigure}%
  \hfill
  \begin{subfigure}{.45\textwidth}
      \includegraphics[width=0.99\textwidth]{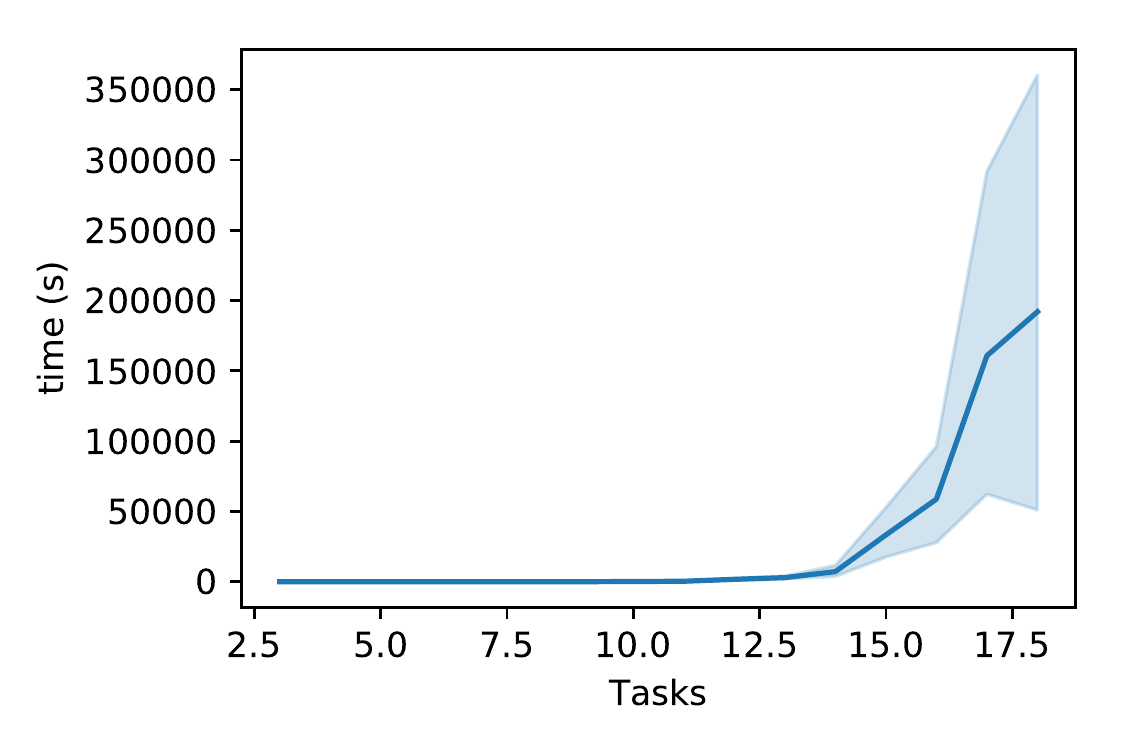}
      \caption{\label{fig:runt-optim-alloc-tasks} CPU time of the optimal allocation algorithm vs.\ the number of components.
        Each experiment with n components was measured 5 times with 5 to 20 nodes.}
    \end{subfigure}
    \caption{\label{fig:runt-optim-alloc} Runtime for optimal allocation.}
\end{figure}

As expected, the optimal allocation algorithm scales very badly (non-polynomially). Figure~\ref{fig:runt-optim-alloc} shows the runtime of the algorithm for varying problem sizes. The shaded area shows the variance with the non-shown parameter (different application sizes for the network node graph, differing network sizes for the application node graph). The time needed for finding the optimal allocation grows unwieldy very quickly.

\begin{figure}
  \centering
  
  \begin{subfigure}{.45\textwidth}
    \centering
    \includegraphics[width=\textwidth]{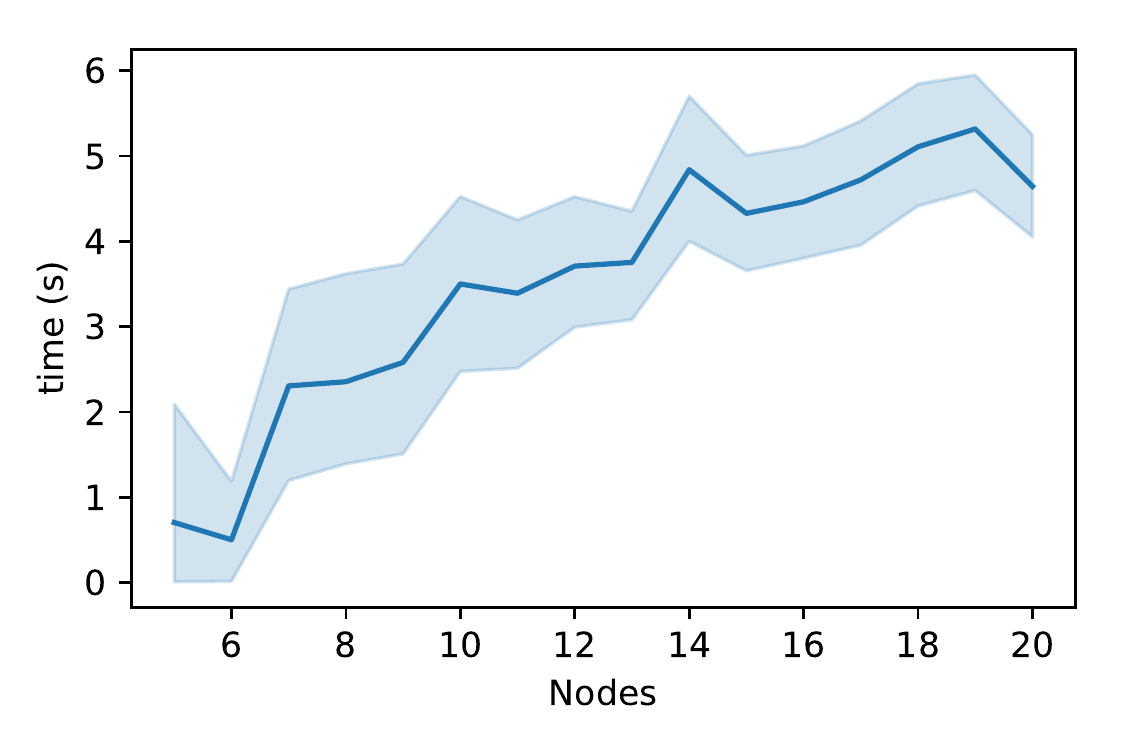}
    \caption{\label{fig:runt-alloc-heur-nodes} CPU time of the allocation heuristic vs.\ the number of nodes. Each experiment with n nodes was measured 5 times with 3 to n-1 components.}
  \end{subfigure}%
  \hfill
  \begin{subfigure}{.45\textwidth}
    \centering
    \includegraphics[width=.99\textwidth]{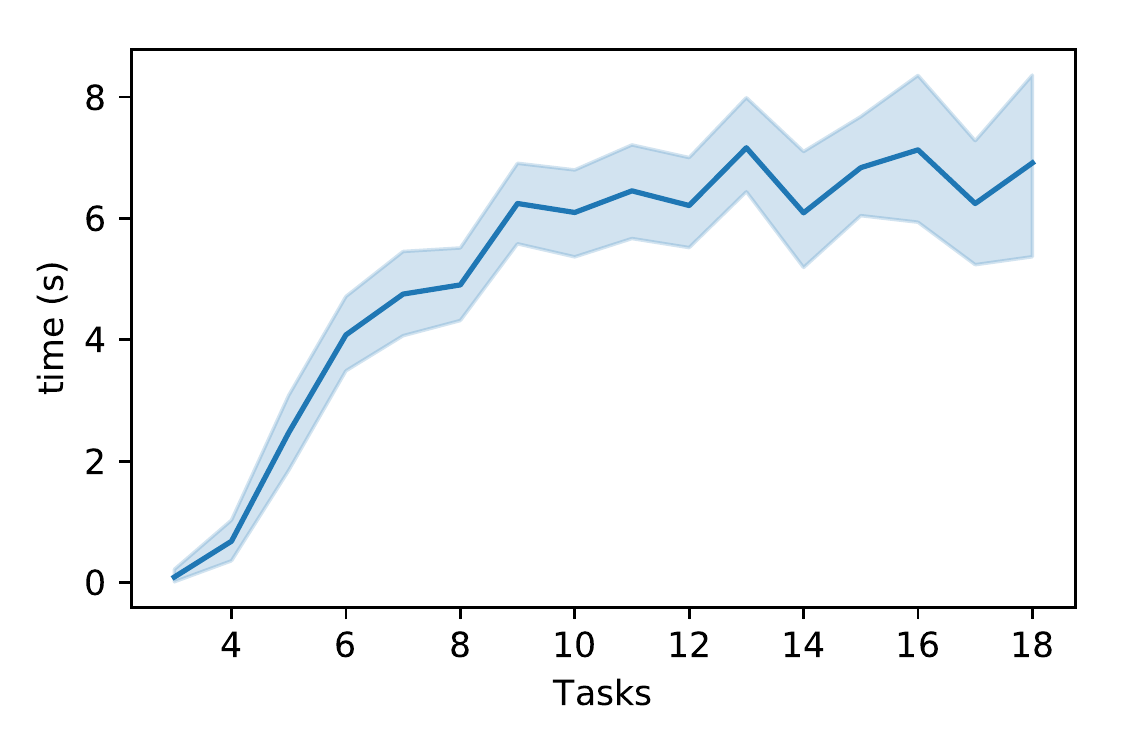}
    \caption{\label{fig:runt-alloc-heur-tasks} CPU time of the allocation heuristic vs.\ the number of components. Each experiment with n components was measured 5 times with 5 to 20 nodes.}
  \end{subfigure}
  \caption{\label{fig:runt-alloc-heur} Heuristic runtime.}
\end{figure}

In comparison, the heuristic presented in equation (\ref{eq:13}) finds a solution much more quickly. Figure~\ref{fig:runt-alloc-heur} shows the runtime of the heuristic for different network and application sizes. For the slowest case for the full allocation, the heuristic takes 8 seconds of CPU time, while the solver consumes 864104 seconds (about 10 days) of CPU time for finding the optimal allocation. The allocation evaluation was executed on an Amazon EC2 \texttt{m4.10xlarge} machine with 40 virtual cores and 160 GiB of memory. Peak memory use was 51 GiB\@. However, the heuristic loses about 30\% of energy efficiency over the optimal algorithm. Specifically, 50\% of the solutions achieve between 60\% and 80\% of the energy efficiency of the optimal case.

\subsection{Energy-efficient Blockchain over Wireless Networks}
\label{sec:enabling_tech-Blockchain}


The last enabling technology is \gls{dlt}, which provides a tamper-proof ledger distributed for the nodes of the \gls{iiote}. The energy and latency cost of implementing \gls{dlt} over wireless links and with constrained \gls{iot} devices is oftentimes overlooked. In general, the latency and energy budgets are highly impacted by the wireless access protocol. 

\subsubsection{System model}

As introduced in~\cite{8977445}, there are two architectural choices for \gls{iot} DLT. The conventional one is to have \gls{iot} devices that receive complete blocks from the Blockchain to which they are connected, and locally verify the validity of the \gls{pow} solution and the contained transactions. This configuration provides the maximum possible level of security. However, this requires high storage, energy and computation resources, since the node needs to store the complete Blockhain and to check all transactions. This makes it infeasible for many \gls{iot} applications. Instead, we consider the second option where the \gls{iot} device is a \emph{light node} that receives only the headers from the Blockchain nodes. These headers contain sufficient information for the \gls{poi}, i.e., to prove the inclusion of a transaction in the block without the need to download the entire block body. Furthermore, the device defines a list of (few) events of interest, such as modifications to the state of a smart contract or transactions from/to a particular address. 

The communication model for this \emph{lightweight} version is as follows. The \gls{iot} devices transmit data to the Blockchain using the edge infrastructure. Specifically, a \gls{nbiot} cell with the base station located in its center is considered, with $N$ devices uniformly distributed within the area. The base station, which is designated as a \emph{full} DLT node connected to the Blockchain, is the DLT-anchor for the \gls{iot} devices. For the radio resource management, we adapt the queueing model of~\cite{energymodel} to our scenario, where the uplink and downlink radio resources are modeled as two servers that visit and serve their respective inter-dependent traffic queues.  

\subsubsection{\gls{e2e} latency}
\gls{nbiot} provides three coverage classes namely normal, extreme, and robust class for serving limited-resource devices and suffering  various pathloss levels\cite{nbiotsmartcity}. Minimum latency and throughput requirements need to be maintained in the extreme coverage class, whereas enhanced performance is ensured in the extended or normal coverage class. Without loss of generality, we consider only normal and extreme coverage class, i.e., the number of classes $C=2$. A class is assigned to a device based on the estimated path loss, with the base station informing the assigned device of the dedicated path between them. Class $j$ and $\forall_j$ are supported by the replicas number $c_j$, which are transmitted based on  data and the control packet \cite{energymodel}. Particularly, the reserved \gls{nprach} period of class $j$ is denoted by $c_j\tau$. The unit length $\tau$ of the \gls{nprach} for the class of coverage is denoted by $c_j = 1$. $t_j$ is the average time interval  between two consecutive scheduling of \gls{nprach} of class $j$, whereas the average time duration between two consecutive \gls{npdcch} occurrences is denoted by $d$. 

The total \gls{e2e} latency includes two parts: (1) the latency $L_{UeD}$ of transmissions of uplink and downlink between \gls{iot} devices and the base station (the wireless communication latency); (2) and the latency $L_{DLT}$ due to the DLT verification process. I.e., $L = L_{UeD} + L_{DLT}$.

The wireless communication latency of \gls{nbiot} uplink and downlink can be formulated as: 
\begin{equation}
        L_{UeD}  = L^{u} + L^{d} = L^{u}_{sync} + L^{u}_{rr} + L^{u}_{tx}  + L^{d}_{sync} +  L^{d}_{rr} + L^{d}_{rx},
\end{equation}
where $L^u_{sync}$, $L^u_{rr}$, $L^u_{tx}$, $L^d_{synch}$,$L^d_{rr}$, and $L^d_{rx}$ are energy consumption of synchronization, resource reservation, and data transmission of uplink and downlink, respectively. $L^u_{sync}$ has been defined in \cite{lsync} with the values of $0.33s$. $L_{rr}$ is given as:%
\begin{equation}
    L_{rr} = \sum_{l=1}^{N_{r_{max}}} (1-P_{rr})^{l-1} P_{rr}l(L_{ra} + L_{rar}),
\end{equation}
in which, $N_{r_{max}}$ is the maximum number of attempts, $P_{rr}$ is the probability of successful resource reservation in an attempt, $L_{ra} = 0.5t + \tau$, is the expected latency in sending an \gls{ra} control message, $\tau$ is the unit length and equal to the \gls{nprach} period for the coverage class 1 which is varied from $40$\,ms to $2.56$\,s \cite{lsync}, and $L_{rar}=0.5d + 0.5 \mathcal{Q} fu +u $, is the expected latency in receiving the RAR message, where $\mathcal{Q}$ are requests waiting to be served. 
%


In the following, we provide a simple technique based on \emph{drift approximation}~\cite{drift_approx} to calculate $P_{rr}$ recursively. Therefore, we treat the mean of the random variables involved in the process as constants. Besides, we assume that sufficient resources are available in the \gls{npdcch} so that failures only occur due to collisions in the \gls{nprach} or to link outages.

Let $\lambda^a=\lambda^u+\lambda^d$ be the arrival rate of access requests per \gls{nprach} period and $\lambda^a(l)$ be the mean number of devices participating in the contention with their $l$-th attempt. Note that in the steady state $\lambda^a(l)$ remains constant for all \gls{nprach} periods. Next, let 
$\lambda^a_{tot}=\sum_{l=1}^{N_{r_{max}}} \lambda^a(l)$. The collision probability in the \gls{nprach} can be calculated using the drift approximation for a given value of $\lambda^a_{tot}$ and for a given number of available preambles $K$ as:
\begin{equation}
    P_\text{collision}(\lambda^a_{tot})=1-\left(1-\frac{1}{K}\right)^{\lambda^a_{tot}-1}\approx 1-e^{-\frac{\lambda^a_{tot}}{K}}.
\end{equation}
\rv{From there, we approximate the probability of resource reservation as a function of $\lambda^a_{tot}$ as
$P_{rr}(\lambda^a_{tot})\approx p_d\,    e^{-\frac{\lambda^a_{tot}}{K}}.$ }
This allows us to define $\lambda^a_{tot}$ as: 
\begin{equation}
\lambda^a_{tot}= \lambda^a +\left(1-P_{rr}(\lambda^a_{tot})\right)\sum_{l=2}^{N_{r_{max}}} \lambda^a(l),
\label{eq:lambdatot}
\end{equation}%
since $\lambda^a(l)=\left(1-P_{rr}(\lambda^a_{tot})\right)\lambda^a(l-1)$ for $l\geq2$ and $\lambda^a(1)=\lambda^a$. Finally, from the initial conditions $\lambda^a(l)=0$ for $l\geq2$, the values of $\lambda^a(l)$ and $\lambda^a_{tot}$ can be calculated recursively by: 1) applying~\eqref{eq:lambdatot}; 2) calculating $P_{rr}(\lambda^a_{tot})$ for the new value of $\lambda^a_{tot}$; and 3) updating the values of $\lambda^a(l)$. This process is repeated until the values of the variables converge to a constant value. The final value of $P_{rr}(\lambda^a_{tot})$ is simply denoted as $P_{rr}$ and used throughout the rest of the paper.

Assuming that the transmission time for the uplink transactions follows a general distribution with the first two moments $l_1, l_2$, the first two moments of the distribution of the packet transmission time are $s_1 = \left(f_1l_1\right)/\left(\mathcal{R}w\right)$, and  $s_2 = \left(f_1l_2\right)/\left(\mathcal{R}^2w^2\right)$. Applying the results from \cite{ltx}, considering $L_{tx}$ as a function of scheduling of NPUSCH, we have: 
\begin{equation}
    L_{tx} = \frac{f\lambda^us_1s_2}{2s_1(1-fGs_1)} + \frac{f\lambda^us_1^2}{2(1-f\lambda^us_1)} + \frac{l_1}{\mathcal{R}^uw},
\end{equation}
where $\mathcal{R}^u$ is the average uplink transmission rate, $\lambda^u = \lambda_s +\lambda_b$, and $f(\lambda_s + \lambda_b)s_1$ is the mean batch-size. The latency of data reception is defined as: 
\begin{equation}
    L_{rx} = \frac{0.5Fh_1t^{-1}}{h_1(1-Fht^{-1})} + \frac{Fh_1}{1-Fht^{-1}} + \frac{m_2}{\mathcal{R}^dy},
\end{equation}
in which, $h_1=fm_1(\mathcal{R}^dy)^{-1}$, $h_2 = fm_2 ((\mathcal{R}^d)^2y^2)^{-1}$  
are two moments of distribution of the packet transmission time, assuming that the packet length follows a general distribution with moments $m_1$, $m_2$, $F=f\lambda^d t$, $\mathcal{R}^d$ is downlink data transmission rate.

Next, we calculate the second latency component, corresponding to the \gls{dlt} verification process. Consider a DLT network that includes $M$ miners. These miners start their Proof-of-Work (PoW) computation at the same time and keep executing the PoW process until one of the miners completes the computational task by finding the desired hash value \cite{bitcoin}. When a miner executes the computational task for the POW of current block, the time period required to complete this PoW can be formulated as an exponential random variable $W$ whose distribution is $f_W(w) = \lambda_c e^{-\lambda_c w}$, in which $\lambda_c= \lambda_0 P_c$ \rv{represents} the computing speed of a miner, $P_c$ is power consumption for computation of a miner, and $\lambda_0$ is a constant scaling factor. Once a miner completes its PoW, it will broadcast messages to other miners, so that other miners can stop their PoW and synchronize the new block. %
\begin{equation}
    L_{tM} = L_{newB} + L_{getB} + L_{transB}
\end{equation}

\rv{In (18),} $L_{newB}$, $L_{getB}$, and $L_{transB}$, are latencies of sending hash of new mined block, requesting new block from neighboring nodes, and new block transmission, respectively. $L_{newB}$ and $L_{transB}$ are computed using uplink transmission, while $L_{getB}$ is computed based on downlink transmission as described in previous section. 

\begin{figure}
    \centering
    \centering
    \begin{subfigure}{0.48\textwidth}
    \includegraphics[width=0.9\linewidth]{ 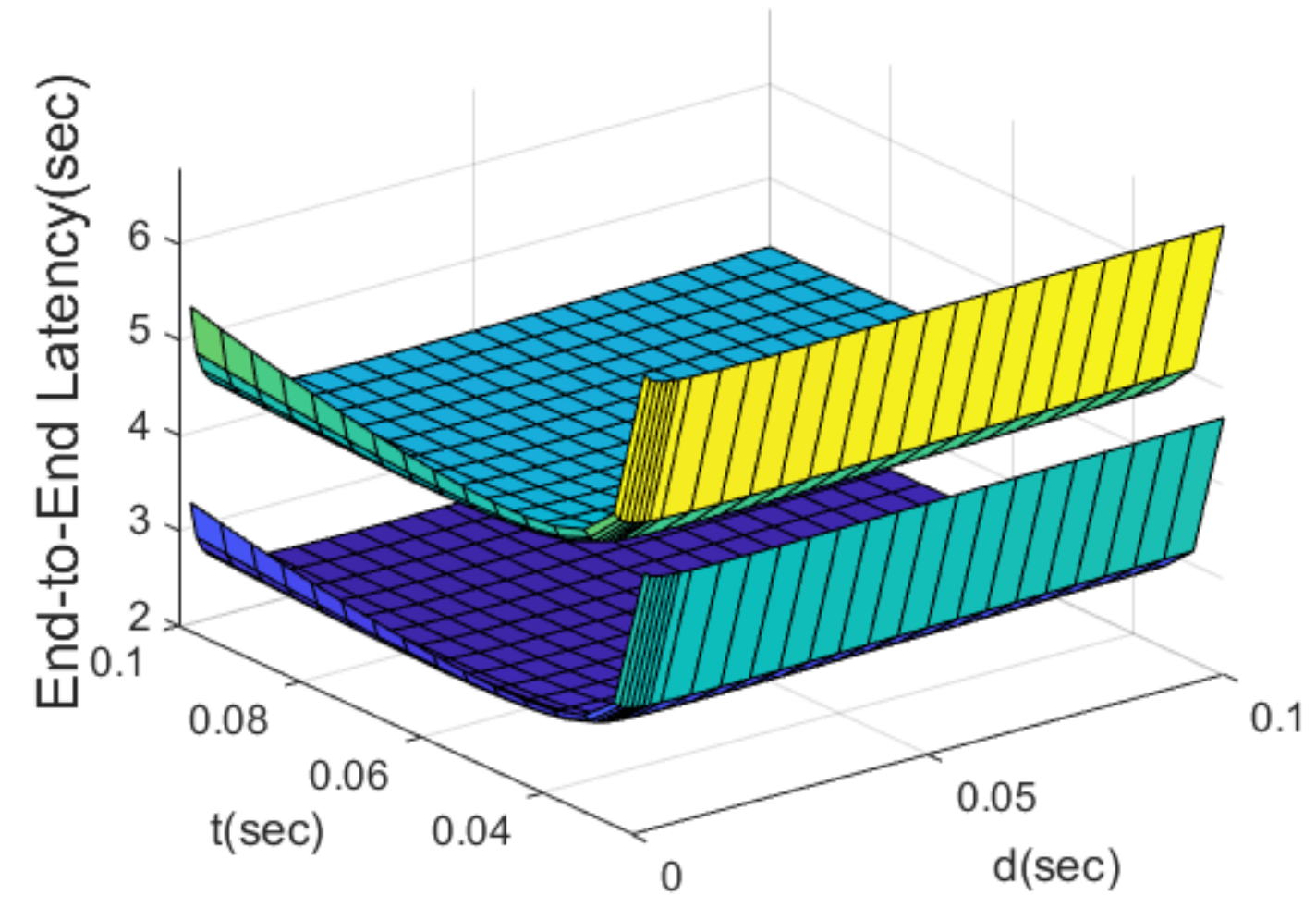} 
    \caption{End-to-End Latency (sec) }
    \label{fig:latency}%
    \end{subfigure}
    \begin{subfigure}{0.48\textwidth}
    \includegraphics[width=0.9\linewidth]{ 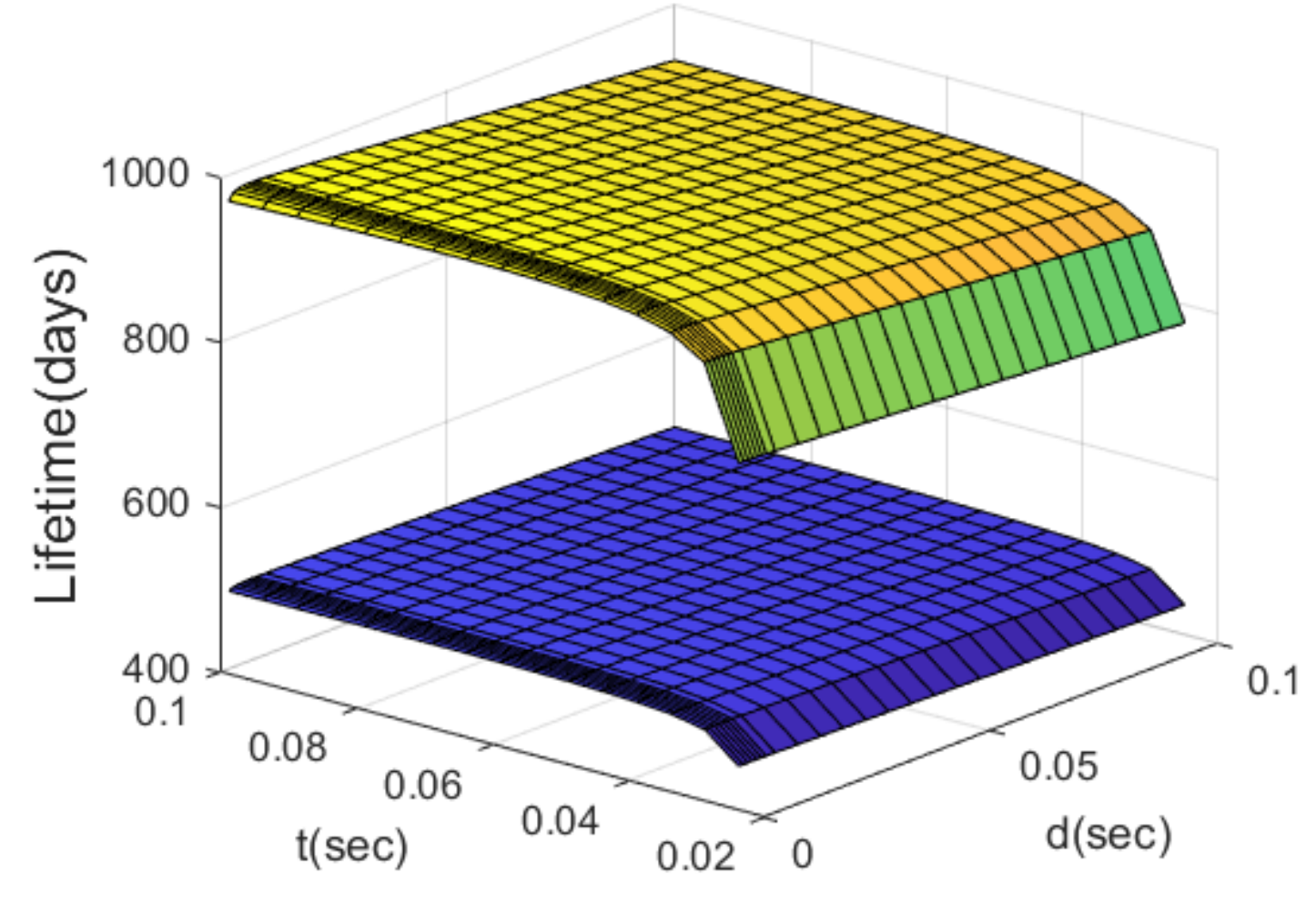}
    \caption{Energy consumption}
    \label{fig:energy}
    \end{subfigure}%
    
    \caption{Latency and Energy Consumption}
    \label{fig:dltperformance}
\end{figure} 

For the PoW computation, a miner $i*$, first finds out the desired PoW hash value, $i*= \min_{i \in M } w_i$. The fastest PoW computation among miners is $W_{i*}$, the complementary cumulative probability distribution of $W_{i*}$ could be computed as $Pr(W_{i*} > x)= Pr(\min_{i \in M} (W_i) > x) = \prod_{i=1}^H Pr(W_i > x) = (1 - Pr(W < x))^M$. Hence, the average computational latency of miner $i*$ is described as: 
%

\begin{align}
    \begin{split}
    L_{W_{i*}} &= \int_0^\infty (1 - Pr(W \leq x))^M \dd x
    \\&= \int_0^\infty e^{-\lambda_cMx}\dd x = \frac{1}{\lambda_cM}
    \end{split}
\end{align}%

The total latency required from DLT verification process is $L_{DLT} = L_{tm} + L_{W_{i*}}$. 


\subsubsection{Energy consumption} Analogously to the latency, the energy consumption is divided in the wireless communication (uplink/downlink) and the \gls{dlt} verification. 

The total energy consumption in the wireless communication is written as follows:
\begin{align}
    \begin{split}
        E_{UD}   &=  E^{u} + E^{d} \\ &= E^{u}_{sync} + E^{u}_{rr} + E^{u}_{tx} +E^{u}_{s} + E^{d}_{sync} +  E^{d}_{rr}  + E^{d}_{rx} + E^{d}_{s},
    \end{split}
\end{align}%
in which, $E^u_{sync}$, $E^u_{rr}$, $E^u_{rr}$, $E^d_{sync}$,$E^d_{rr}$, and $E^d_{rx}$ are energy consumption of synchronization, resource reservation, and data transmission of uplink and downlink, respectively. Each of them are formally defined as follows:
\begin{align}
 & E_{sync} = P_l \cdot L_{sync} \\
 & E_{rar} = P_l \cdot L_{rar} \\
 &  E_{rr} = \sum_{l=1}^{N_{max}} (1-P_{rr})^{l-1}  \cdot P_{rr}  \cdot (E_{ra} + E_{rar}) \\
 & E_{ra} = (L_{ra} - \tau) \cdot P_I + \tau \cdot (P_c + P_e P_t) \\
 & E_{tx} = (L_{tx} - \frac{l_a}{\mathcal{R}^uw}) \cdot P_I + (P_c + P_e P_t)\frac{l_a}{\mathcal{R}^uw} \\
 & E_{rx} = (L_{rx} - \frac{m_1}{\mathcal{R}^dy}) \cdot P_I + P_l \frac{m_1}{\mathcal{R}^dy}
\end{align}
where $P_e$, $P_I$, $P_c$, $P_l$, and $P_t$ are the power amplifier efficiency, idle power consumption, circuit power consumption of transmission, listening power consumption, and transmit power consumption, respectively. %

Following the \gls{pow} described above, the average energy consumption of DLT to finish a single \gls{pow} round is:
\begin{equation}
    E_{DLT} = P_c L_{W_{i*}} + P_{t} L_{tm}
\end{equation}

\subsubsection{Results}
The performance of DLT-based \gls{nbiot} system is shown in Fig. \ref{fig:dltperformance}. The experiments demonstrate the total latency Fig. \ref{fig:latency} and energy efficient Fig. \ref{fig:energy} of a DLT-based \gls{nbiot} system, respectively. 
In Fig. \ref{fig:latency}, the E2E latency is defined as the time elapsed from the generation of a transaction at the \gls{nbiot} device until its verification. 
This includes the latency at the \gls{nbiot} radio link and at the DLT, which comprises the execution time of the smart contract and transaction verification.
We observe that increasing $t$ and $d$ values at the first increases lifetime and decrease latency due to more resources for NPUSCH and NPDSCH, but after certain point increasing $t$ and $d$ decreases the lifetime by increasing the expected time for resource reservation. 
In comparison with the standard \gls{nbiot} system in \cite{9141220, 8536384}, the DLT-based system introduces a slight latency because of addition time of consensus process and transaction verification. This is a latency and security trade-off between standard \gls{nbiot} and DLT-based systems.


%% file: 5-towards-energy-efficient-intelligent-iot.tex
\begin{figure*}[ht!]
    \centering
    \includegraphics[width=0.8\linewidth]{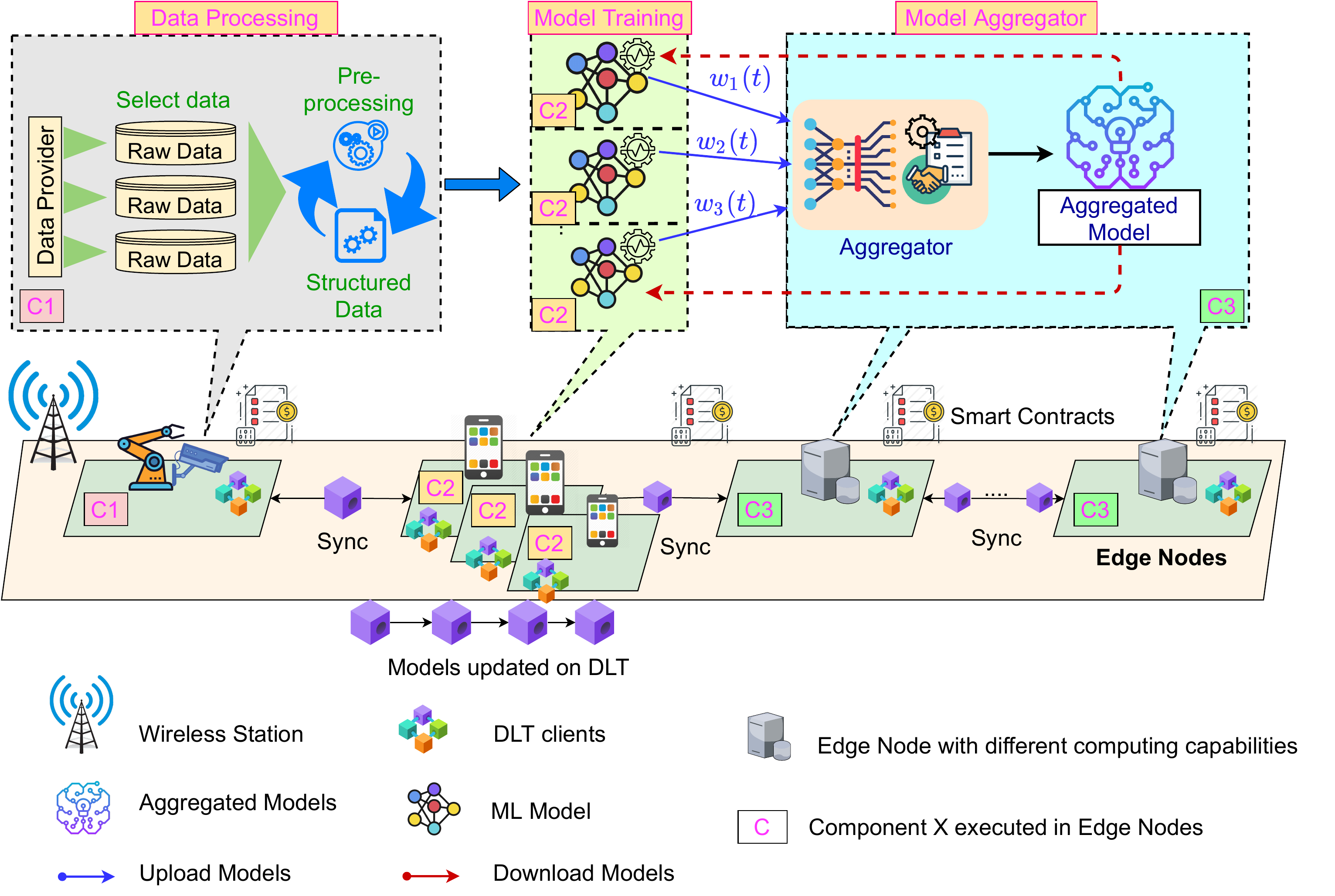}
    \caption{Integration of enabling technologies.}
    \label{fig:integration}
\end{figure*}

\section{Towards energy-efficient intelligent IoT environments} \label{sec:discussion}

Having the energy-performance characterization for each of the enabling technologies (Section \ref{sec:enabling_tech}), we describe next how they interact with each other in \gls{iiote}. For this, we consider the scenario in Figure~\ref{fig:integration}, where a given learning application is considered. 
We split the task into sub-tasks such as data processing, data training and model aggregation and distribute them in a decentralized way. Each of these sub-tasks (Section~\ref{sec:enabling_tech-DL}) constitute the application components ($C_1, C_2, ...$) that can be run at the available edge nodes. 
The optimal allocation of sub-tasks to edge nodes is determined using the ILP-based algorithms presented in Section~\ref{sec:enabling_tech-Optimizing}.
The required trustworthiness (i.e., assuring security, privacy, immutability and transparency) between sub-tasks is provided through DLT (Section \ref{sec:enabling_tech-Blockchain}).
The heterogeneity of devices, capabilities and tasks is exploited accordingly:
The edge servers with high computation capability are selected to operate the DLT activities, e.g, block mining, and aggregate the ML models (the head workers if the learning paradigms in Section~\ref{sec:enabling_tech-DL} are applied), while more constrained edge devices or mobile devices are setup as DLT light clients that can participate in local training (the tail workers) and consensus. The involved network components can communicate via wireless long-range communication \gls{nbiot} channels.

In detail, the communication workflow of the proposed scheme can be summarized as follows: 

\begin{itemize}
    \item \textbf{Step 1}: The data processing can be completed in different edge devices with limited resources. The selected data from the data provider is pre-processed and structured. This process includes both a data engineering and feature engineering sub-process, in which data engineering converts the raw data into prepared data and feature engineering tunes the prepared data to create features expected by ML models. 
    \item \textbf{Step 2}: Then, the edge nodes or IoT devices, which are responsible for training, compute the local model based on its own private data and then publish the local model to its associated edge server via, e.g., NB-IoT by registering with active smart contracts to upload their result securely until the results are incorporated in the final aggregation and generation of DLT transactions. 
    \item \textbf{Step 3}: Next, the edge servers with ML aggregation responsibility gather transactions and arrange them in blocks following the Merkle tree. The structure of a DLT involves the hash of the previous block, a timestamp, 'nonce' and the structure of hash tree. These edge servers with high computational capacity join in the DLT mining process to verify the created blocks and operate consensus in the edge network. After completing the mining process, the verified blocks are added to the ledger, and synchronized among the nodes. The local models are published in the distributed ledger. Hence, the powerful edge servers can compute the global model directly based on the aggregation rules defined in smart contracts. 
\end{itemize}

The advantages of this integration are two-folds. First,  by distributing the tasks to different edge nodes with different computing capacities, the IoT devices or edge nodes with limited resources can save significant amount of energy required for training or mining and they can achieve lower latency. 
Second, by leveraging the DLT, the updates of ML models are securely formed in encrypted transactions and hashed blocks, which significantly enhances the security and privacy of distributed learning in the edge networks. The DLT provides trust, transparency and immutability baseline for distributed learning to guarantee the security and privacy of data and ML models, and naturally addresses the single-point of failure problem of the current standard \gls{fl} approach 
 that relies on a centralized server to aggregate the models. 
Although the integration of enabling technologies introduces advantages, it also has some drawbacks, for example, the time required of DLT mining will increase the total latency of the system. This is a trade-off between trust and communication latency which we discussed in \cite{danzi2020communication},\cite{nguyen2020trusted}.

%% file: 6-conclusions.tex
\section{Conclusions and Future Work} \label{sec:conclusions}

In this paper, we address the evolution of next-generation of \gls{iot} networks towards the edge, driven by the introduced intelligent \gls{iot} environments. We use the \gls{iiote} as the basic building block to characterize the tradeoff energy-performance of the three key enabling technologies, learning, edge computing and distributed ledger. 
Edge intelligence must rely on distributed paradigms such as \gls{fl}, and we have shown how exploiting spatial and temporal sparsity and quantization 
can significantly improve the performance and reduce the energy consumption. 
Moreover, we have discussed the distribution of the \gls{fl} model aggregator and the rest of sub-tasks to make the framework more robust against failures. 
For edge computing, the optimal allocation of the application components to network resources is important to efficiently use the available infrastructure and optimize its energy consumption. 
\gls{dlt} is a flexible solution for trustworthiness in these environments, but the energy and latency cost of implementing \gls{dlt} over wireless and constrained devices is oftentimes overlooked. We have analyzed these parameters using \gls{nbiot} as the baseline wireless technology. 

In the integration of these technologies in  \gls{iiote}, we have shown the interactions among them, which provides the basis towards an energy model and evaluation that encompasses the contribution of each element. For instance, the learning and computation models can be easily broaden to consider the allocation of the different sub-tasks of the learning application in a representative topology, with each learning action and resource allocation playing the role of an action to be recorded in the \gls{dlt}. 
Future work also includes extending the proposed solutions to dynamic environments where agents move and edge nodes are not always available. This is already supported by the presented dynamic \emph{head/tail} learning paradigms but the integration of a dynamic resource allocation and \gls{dlt} framework is pending. Another necessary direction is to investigate the joint optimization of the computing and communication resources from the energy perspective. 